\documentclass[letterpaper,11pt]{article}
\pdfoutput=1

\usepackage[utf8]{inputenc}

\usepackage{jheppub}
\usepackage{multirow, amsthm}
\usepackage{braket}
\usepackage{mathrsfs}
\usepackage{tikz}
\usepackage[caption=false]{subfig}
\usepackage{xspace}
\usepackage{xcolor}
\usepackage{float,color}
\usepackage{siunitx}
\usepackage[ruled,norelsize]{algorithm2e}
 \usepackage{comment}

\usepackage{accents}
\newlength{\dhatheight}
\newcommand{\doublehat}[1]{%
    \settoheight{\dhatheight}{\ensuremath{\hat{#1}}}%
    \addtolength{\dhatheight}{-0.35ex}%
    \hat{\vphantom{\rule{1pt}{\dhatheight}}%
    \smash{\hat{#1}}}}

\definecolor{darkred}{rgb}{0.5,0.0,0.0}
\definecolor{darkblue}{rgb}{0.0,0.0,0.9}
\definecolor{darkerblue}{rgb}{0.0,0.0,0.5}
\definecolor{darkgreen}{rgb}{0.0,0.5,0.0}
\definecolor{black}{rgb}{0.0,0.0,0.0}
\definecolor{brown}{rgb}{0.6,0.4,0.2}

\title{\boldmath Extending the Bump Hunt with Machine Learning}
\author[1,2]{Jack H. Collins}
\author[3]{Kiel Howe}
\author[4,5]{and Benjamin Nachman}
\affiliation[1]{\normalsize\it Maryland Center for Fundamental Physics, Department of Physics, University of Maryland, College Park, MD 20742, USA}
\affiliation[2]{\normalsize\it Department of Physics and Astronomy, Johns Hopkins University, Baltimore, MD 21218, USA}
\affiliation[3]{\normalsize\it Fermi National Accelerator Laboratory, Batavia IL 60510, USA}
\affiliation[4]{\normalsize\it Physics Division, Lawrence Berkeley National Laboratory, Berkeley, CA 94720, USA}
\affiliation[5]{\normalsize\it Simons Institute for the Theory of Computing, University of California, Berkeley, Berkeley, CA 94720, USA}

\emailAdd{jhc296@umd.edu}
\emailAdd{khowe@fnal.gov}
\emailAdd{bpnachman@lbl.gov}

\abstract{
The oldest and most robust technique to search for new particles is to look for `bumps' in invariant mass spectra over smoothly falling backgrounds.  We present a new extension of the bump hunt that naturally benefits from modern machine learning algorithms while remaining model-agnostic.  This approach is based on the Classification Without Labels (CWoLa) method where the invariant mass is used to create two potentially mixed samples, one with little or no signal and one with a potential resonance.  Additional features that are uncorrelated with the invariant mass can be used for training the classifier.  Given the lack of new physics signals at the Large Hadron Collider (LHC), such model-agnostic approaches are critical for ensuring full coverage to fully exploit the rich datasets from the LHC experiments.  In addition to illustrating how the new method works in simple test cases, we demonstrate the power of the extended bump hunt on a realistic all-hadronic resonance search in a channel that would not be covered with existing techniques.
}

\begin{document} 
\maketitle
\flushbottom

\section{Introduction}
\label{sec:intro}


Searching for new resonances as bumps in the invariant mass spectrum of the new particle decay products is one of the oldest and most robust techniques in particle physics, from the $\rho$ meson discovery~\cite{PhysRev.126.1858} and earlier up through the recent Higgs boson discovery~\cite{Aad:2012tfa,Chatrchyan:2012xdj}.  This technique is very powerful because sharp structures in invariant mass spectra are not common in background processes, which tend to produce smooth distributions.  As a result, the background can be estimated directly from data by fitting a shape in a region away from the resonance (sideband) and then extrapolating to the signal region.  It is often the case that the potential resonance mass is not known a priori and a technique like the BumpHunter~\cite{Choudalakis:2011qn} is used to scan the invariant mass distribution for a resonance.  In some cases, the objects used to construct the invariant mass (e.g. jet substructure) and their surroundings (e.g. presence of additional forward jets) have properties that can be used to increase the signal purity.  Both ATLAS and CMS\footnote{The references here are the Run 2 results; Run 1 results can be found within the cited papers.  The techniques described in this paper also apply to leptonic or photonic final states, but jets are used as a prototypical example due to their inherent complex structure.} have conducted extensive searches for resonances decaying into jets originating from generic quarks and gluons~\cite{Aaboud:2017yvp,Sirunyan:2016iap,Khachatryan:2015dcf}, from boosted $W$~\cite{Sirunyan:2017acf,Aaboud:2017eta}, $Z$~\cite{Sirunyan:2016wqt}, $Z'$~\cite{Sirunyan:2017dnz,Sirunyan:2017nvi,Aaboud:2018zba} or Higgs bosons~\cite{Sirunyan:2017dgc,Sirunyan:2017isc,Aaboud:2017ahz}, from $b$-quarks~\cite{Aaboud:2016nbq}, as well as from boosted top quarks~\cite{Sirunyan:2017uhk,Aaboud:2018juj}.  There is some overlapping sensitivity in these searches, but in general the sensitivity is greatly diminished away from the target process (see e.g.~\cite{Aguilar-Saavedra:2018xpl,Aguilar-Saavedra:2017zuc,boosted_diboson} for examples).  It is not feasible to perform a dedicated analysis for every possible topology and so some signals may be missed.  Global searches for new physics have been performed by the the LHC experiments and their predecesors, but only utilize simple objects and rely heavily on simulation for background estimation~\cite{ATLAS-CONF-2017-001,CMS-PAS-EXO-10-021,ATLAS-CONF-2012-107,ATLAS-CONF-2014-006,Aktas:2004pz,Aaron:2008aa,Abbott:2000fb,Abbott:2000gx,Aaltonen:2007dg,Aaltonen:2008vt,sleuth,Knuteson:2004nj}.

The tagging techniques used to isolate different jet types have increased in sophistication with the advent of modern machine learning classifiers~\cite{Larkoski:2017jix,Cogan:2014oua,Almeida:2015jua,deOliveira:2015xxd,Baldi:2016fql,Barnard:2016qma,Kasieczka:2017nvn,Butter:2017cot,Komiske:2016rsd,Louppe:2017ipp,ATLAS-CONF-2017-064,ATL-PHYS-PUB-2017-013,ATL-PHYS-PUB-2017-004,CMS-DP-2017-005,CMS-DP-2017-013,ATL-PHYS-PUB-2017-003,Pearkes:2017hku,Datta:2017rhs,Datta:2017lxt,ATL-PHYS-PUB-2017-017,CMS-DP-2017-027,Fraser:2018ieu,Andreassen:2018apy,Macaluso:2018tck}.  These new algorithms can use all of the available information to achieve optimal classification performance and could significantly improve the power of hadronic resonance searches.  Deep learning techniques are able to outperform traditional methods by exploiting subtle correlations in the radiation pattern inside jets.  These correlations are not well-modeled in general~\cite{Barnard:2016qma} which renders classifiers sub-optimal when training on simulation and testing on data.   This is already apparent for existing multivariate classifiers where post-hoc mis-modeling corrections can be large~\cite{Aad:2015ydr,Chatrchyan:2012jua,Aad:2014gea,CMS:2013kfa,CMS-DP-2016-070,Aad:2015rpa,Khachatryan:2014vla,Aad:2016pux,CMS:2014fya}.   Ideally, one would learn directly from data (if possible) and/or combine with other approaches to mitigate potential mis-modeling effects during training (e.g. with adversaries~\cite{Louppe:2016ylz}).

We propose a new method that combines resonance searches with recently proposed techniques for learning directly from data~\cite{Dery:2017fap,Metodiev:2017vrx,Komiske:2018oaa,Cohen:2017exh}.  Simply stated, the new algorithm trains a fully supervised classifier to distinguish a signal region from a mass sideband using auxiliary observables which are decorrelated from the resonance variable under the background-only hypothesis.  A bump hunt is then performed on the mass distribution after applying a threshold on the classifier output.  This is Classification Without Labels (CWoLa)~\cite{Metodiev:2017vrx} where the two mixed samples are the signal region and sideband and the signal is a potential new resonance and the background is the Standard Model continuum.  The algorithm naturally inherits the property of CWoLa that it is fully based on data and thus is insensitive to simulation mis-modeling\footnote{The algorithm also inherits the assumptions of the CWoLa method. In this context, the main assumption will be that signal region and sideband region can only be distinguished with the mass. More details on this are in the next sections. }.  The key difference with respect to Ref.~\cite{Metodiev:2017vrx,Komiske:2018oaa} is that the signal process need not be known a priori.  Therefore, we can become sensitive to new signatures for which we did not think to construct dedicated searches.

In addition to CWoLa, the extended bump hunt shares some features with the sPlot technique~\cite{Pivk:2004ty}.  Our proposed extension to the bump hunt makes use of auxiliary features to enhance the presence of signal events over background events in a target distribution, where the signal is expected to be resonant. Similarly, sPlot provides a procedure for using auxiliary features (`discriminating variables' in the language of Ref.~\cite{Pivk:2004ty}) to extract the distribution of signal and background events in a target distribution (`control variable' in Ref.~\cite{Pivk:2004ty}).  In both cases, the auxiliary features must be uncorrelated with the target feature.  One main difference between the methods is that the extended bump hunt uses machine learning to identify regions of phase space that are signal-like.  A second key distinction between methods is that sPlot takes the distribution of the auxiliary features as input, whereas this information is not required for the extended bump hunt.  

This paper is organized as follows.  Section~\ref{sec:cwolahunting} formally introduces the CWoLa hunting approach and briefly discusses how auxiliary information can be useful for bump hunting.  Then, Sec.~\ref{sec:example} uses a simplified example to show how a neural network can be used to identify new physics structures from pseudodata.  A complete procedure for applying the CWoLa hunting approach is given in Sec.~\ref{sec:fullmethod}.  Finally, a realistic example based on a hadronic resonance search is presented in Sec.~\ref{sec:physicsexample}.  Conclusions and future outlook are presented in Sec.~\ref{sec:conc}.

\section{Bump Hunting using Classification Without Labels}
\label{sec:cwolahunting}

In a typical resonance search, events have at least two objects whose four-vectors are used to construct an invariant mass spectrum.  The structure of these objects as well as other information in the event may be useful for distinguishing signal from background even though there may be no other resonance structures.  Let $m_\text{res}$ be a random variable that represents the invariant mass.  The distribution of $m_\text{res}$ given $\text{background}$ is smooth while $m_\text{res}$ given $\text{signal}$ is expected to be localized near some $m_0$.  Let $Y$ be another random variable that represents all other information available in the events of interest.  Define two sets of events: 

\begin{align}
M_1&=\{(m_\text{res},Y)||m_\text{res}-m_0|<\delta\} \hspace{3mm}\text{(the signal region) }\\
 M_2&=\{(m_\text{res},Y)|\delta<|m_\text{res}-m_0|<\epsilon\}\hspace{3mm}\text{(the sideband region)},
 \end{align}
 
\noindent where $\epsilon > \delta$.  The value of $\delta$ is chosen such that $M_1$ should have much more signal than $M_2$ and the value of $\epsilon$ is chosen such that the distribution of $Y$ is nearly the same between $M_1$ and $M_2$.  CWoLa hunting entails training a classifier to distinguish $M_1$ from $M_2$ using $Y$ and then performing a usual bump hunt on $m_\text{res}$ after placing a threshold on the classifier output.  This procedure is then repeated for all mass hypotheses $m_0$.  Note that nothing is assumed about the distribution of $Y$ other than that it should be nearly the same for $M_1$ and $M_2$ under the background-only hypothesis.  

Ideally, $Y$ incorporates as much information as possible about the properties of the objects used to construct the invariant mass and their surroundings.  The subsequent sections will show how this can be achieved with neural networks.  To build intuition for the power of auxiliary information, the rest of this section provides analytic scaling results for a simplified bump hunt with the most basic case: $Y\in\{0,1\}$.  

Suppose that we have two mass bins $M_1$ and $M_2$ and the number of expected events in each mass bin is $N_b$.  Further suppose that the signal is in at most one of the $M_i$ (not required in general) and the expected number of signal events is $N_s$.  A version of the bump hunt would be to compare the number of events in $M_1$ and $M_2$ to see if they are significantly different.  As a Bernoulli random variable, $Y$ is uniquely specified by $\Pr(Y=1)$.  Define $\Pr(Y=1|\text{background})=p$ and $\Pr(Y=1|\text{signal})=q$.  The purpose of CWoLa hunting is to incorporate the information about $Y$ into the bump hunt.  By only considering events with $Y=1$, the significance of the signal scales as $qN_s/\sqrt{N_bp}$. Therefore, the information about $Y$ is useful when $q>\sqrt{p}$.  

\begin{figure}[h!]
\centering
\includegraphics[width=0.5\textwidth]{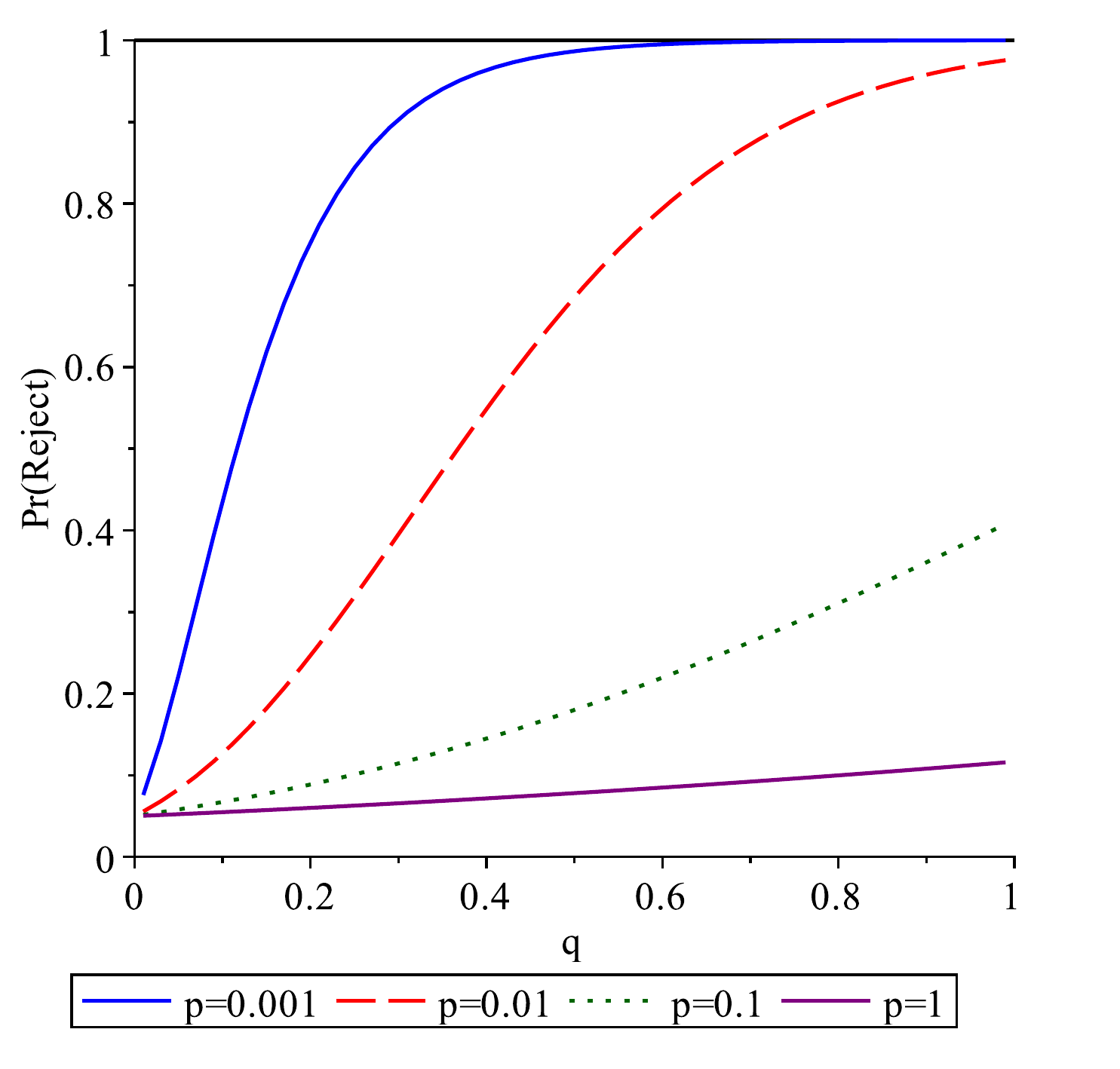}
\label{fig:analytic}
\caption{The probability to reject the SM as a function of $q$ for fixed values of $p$ as indicated in the legend when only considering events with $Y=1$.  The expected background is fixed at 1000 and the expected BSM is fixed at 20.  When $q=0$, there is no signal and therefore the rejection probability is 5\%, by construction.  When $q=p=1$, $Y$ is not useful, but the probability to reject is above 5\% simply because there is an excess of events inclusively. }
\end{figure}

More quantitatively, suppose that we declare discovery of new physics when the number of events with $Y=1$ in $M_1$ exceeds the number of events with $Y=1$ in $M_2$ by some amount.  Under the background-only case, for $N_b\gg 1$, the difference between the number of events in $M_1$ and $M_2$ with $Y=1$ is approximately normally distributed with mean $0$ and variance $2N_bp$.  If we want the probability for a false positive to be less than 5\%, then the threshold value is simply $\sqrt{2N_bp}\times\Phi^{-1}(0.95)$, where $\Phi$ is the cumulative distribution function of a standard normal distribution.  Ideally, we would like to reject the SM often when there is BSM, $N_s>0$.  Figure~\ref{fig:analytic} shows the probability to reject the SM for a one-bin search using $N_b=1000$ and $N_s=20$ for different values of $p$ as a function of $q$.  The case $p=q=1$ corresponds to the standard search that does not gain from having additional information.  However, away from this case, there can be a significant gain from using $Y$, especially when $p$ is small and $q$ is close to $1$.  In the case where $Y$ is a \textit{truth bit}, i.e. $p=1-q=0$, the SM is rejected as long as a single BSM event is observed.  By construction, when $q\rightarrow 0$ (for $p>0$), the rejection probability is 0.05.  Note that when $q<p$, only considering events with $Y=1$ is sub-optimal - this is a feature that is corrected in the full CWoLa hunting approach.

While the model used here is simple, it captures the key promise of CWoLa hunting that will be expanded upon in more detail in the next sections.  In particular, the main questions to address are: how to find $Y$ and how to use the information about $Y$ once it is identified.

\section{Illustrative Example: Learning to Find Auxiliary Information}
\label{sec:example}

This section shows how to identify the useful attributes of the auxiliary information using a neural network.   The example used here is closer to a realistic case, but is still simplified for illustration.  Let the auxiliary information $Y=(x,y)$ be two-dimensional and assume that $Y$ and the invariant mass are independent given the process (signal or background).  This auxiliary information will become the jet substructure observables in the next section.   For simplicity, for each process $Y$ is considered to be uniformly distributed on a square of side length $\ell$ centered at the origin.  The background has $\ell=1$ $(-0.5<x<0.5,-0.5<y<0.5)$ and the signal follows $\ell=w$ $(-w/2 < x < w/2, -w/2 < y < w/2)$. Similarly to the full case, suppose that there are three bins of mass for a given mass hypothesis: a signal region $m_0\pm\Delta$ and mass sidebands $(m_0-2\Delta, m_0-\Delta)$, $(m_0+\Delta, m_0+2\Delta)$.  As in the last section, the signal is assumed to only be present in one bin (the signal region) with $N_s$ expected events.  There are $N_b$ expected background events in the signal region and $N_b/2$ expected events in each of the mass sidebands.  

\begin{figure}[h!]
\centering
\includegraphics[width=0.8\textwidth]{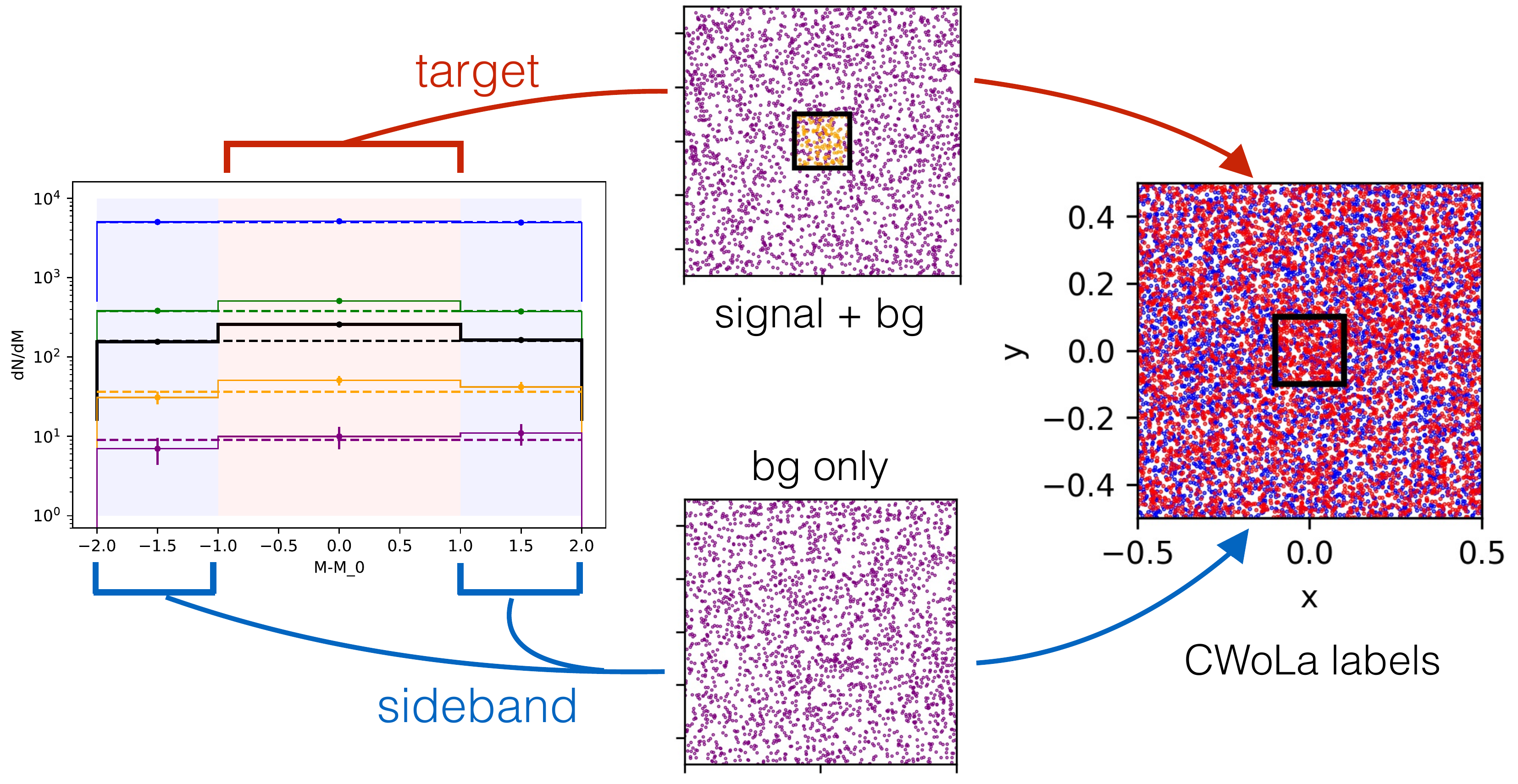}
\caption{An illustration of the CWoLa procedure for the simple two-dimensional uniform example presented in Sec.~\ref{sec:example}.  The left plot shows the mass distribution for the three mass bins, which is uniform for the background.  The blue line is the total number of events and the other lines represent thresholds on various neural networks described in the text leading up to Fig.~\ref{fig:toy_cuts}. The center plots show the $(x,y)$ distribution for the events in each mass bin with truth labels (purple for background and yellow for signal). The black square is the true signal region for this example model, with signal distributed uniformly inside. The right plot shows the combined distribution in the $(x,y)$ plane with CWoLa labels that can be used to train a classifier even without any truth-level information (red for target window, blue sideband). \label{fig:toy_cwola}}
\end{figure}

\begin{figure}[h!]
\centering
\includegraphics[width=0.8\textwidth]{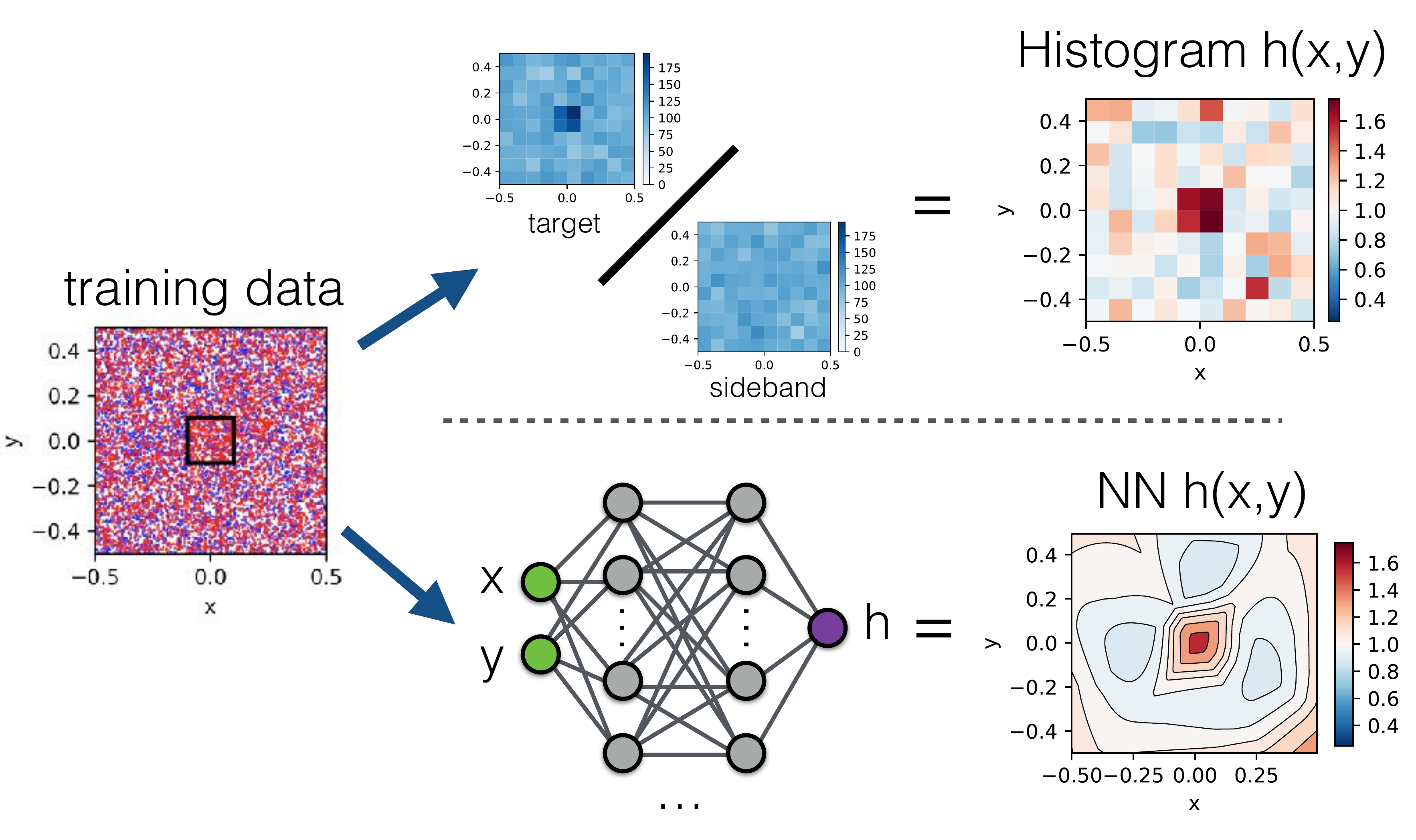}
\caption{The CWoLa-labeled data can be used to construct an estimate for the optimal classifier $h(x,y)=\frac{p_b(x,y) + p_s(x,y)}{p_b(x,y)}$. The top path shows an estimate constructed by histogramming the observed training events in the $(x,y)$ plane. The bottom path shows an estimate constructed by using a neural network trained as described in the text, which can be efficiently generalized to higher dimensional distributions.  The optimal classifier would be 1 outside of the small box centered at the origin and 1.75 inside the box. \label{fig:toy_likelihood_estimators}}
\end{figure}

The model setup described above and used for the rest of this section is depicted in Fig.~\ref{fig:toy_cwola}.  The numerical examples presented below use $N_b=10,000$, $N_s=300$, and $w=0.2$.  Without using $Y$, these values correspond to $N_s/\sqrt{N_b}=3\sigma$.  The ideal tagger (one that is optimal by the Neyman-Pearson lemma~\cite{Neyman289}) should reject all events outside of the square in the $(x,y)$ plane centered at zero with side length $w$.  For the $N_s$ and $N_b$ used here, the expected significance of the ideal tagger is $15\sigma$.  The goal of this section is to show that without using any truth information, the CWoLa approach can recover much of the discriminating power from a neural network trained in the $(x,y)$ plane.  Note that optimal classifier is simply given by thresholding the likelihood ratio~\cite{Neyman289} $p_s(Y)/p_b(Y)$; in this two-dimensional case it is possible to provide an accurate approximation to this classifier without neural networks.  However, these approximations often do not scale well with the dimensionality and will thus be less useful for the realistic example presented in the next section.  This is illustrated in the context of the CWoLa hunting in Fig.~\ref{fig:toy_likelihood_estimators}.


To perform CWoLa hunting, a neural network is trained on $(x,y)$ values to distinguish events in the mass sidebands from the signal region.  Due to the simple nature of the example, it is also possible to easily visualize what the network is learning.  A fully-connected feed-forward network is trained using the \textsc{Python} deep learning library \textsc{Keras}~\cite{keras} with a \textsc{Tensorflow}~\cite{tensorflow} backend.  The network has three hidden layers with (256, 256, 64) nodes. The network was trained with the categorical cross-entropy loss function using the \textsc{Adam} algorithm~\cite{adam} with a learning rate of 0.003 and a batch size of 1024.  The data are split into three equal sets, one used for training, one for validation, and one for testing. The training is terminated based on the efficiency of the signal region cut on the validation data at a fixed false-positive-rate of $2\%$ for the sideband data. If it fails to improve for 60 epochs, the training is halted and the network reverts to the last epoch for which there was a training improvement. This simple scheme is robust against enhancing statistical fluctuations but reduces the number of events used for the final search by a factor of three as only the classifier output on the test set is used for the bump hunt.  In the physical example described later, a more complicated scheme maximizes the statistical power of the available data.  

\begin{figure}[h!]
\centering
\includegraphics[width=0.6\textwidth]{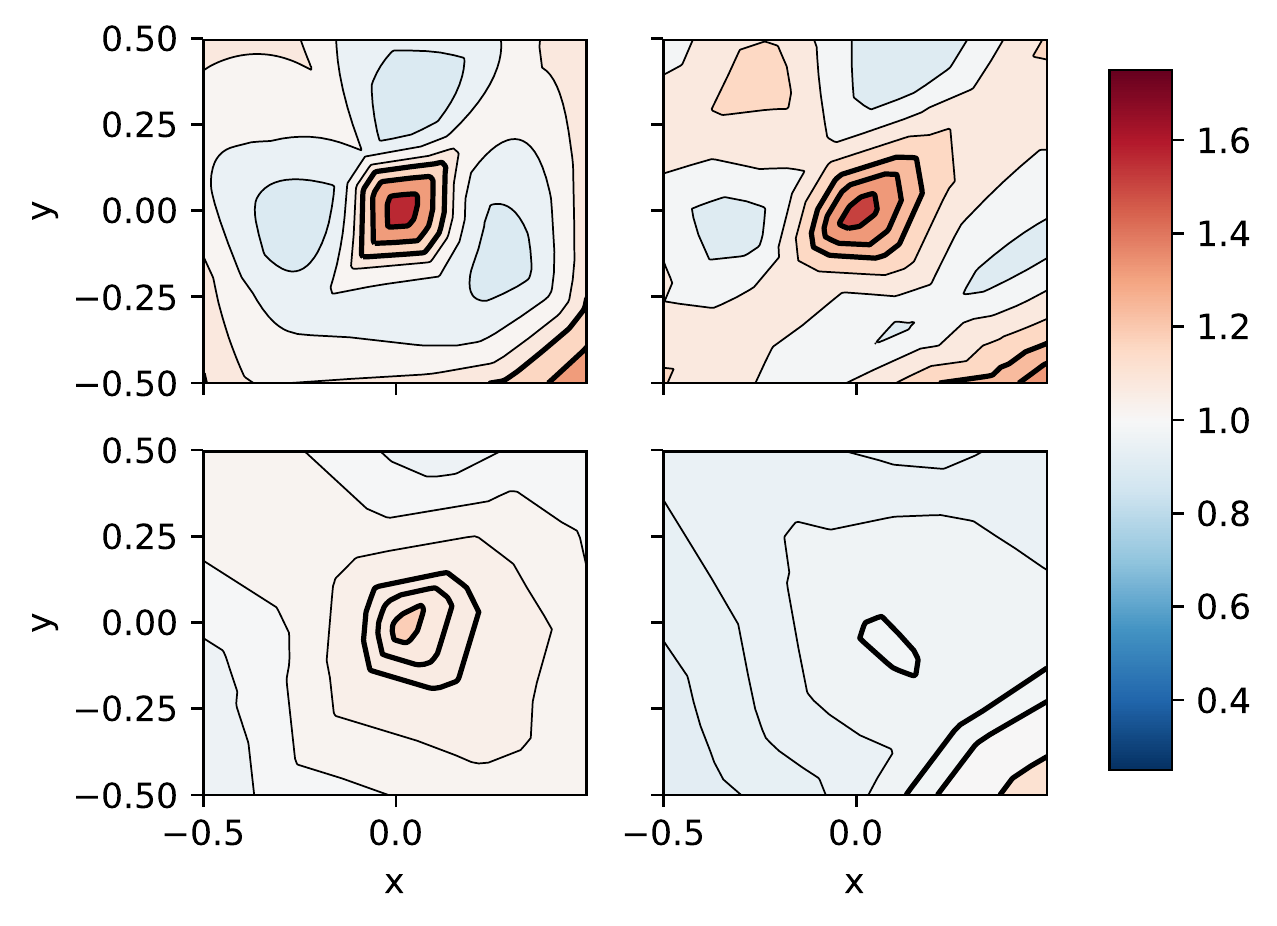}
\caption{The classifier $h(x,y)$ constructed from four independent training runs on the same example two-dimensional model dataset described in the text. The thick contours represent the cuts that would reduce the events in the test data target window by a factor of $\epsilon_{\rm test} = 10\%, 5\%, 1\%$. \label{fig:toy_NNactivations}}
\end{figure}

Visualizations of the neural network trained as described above are presented in Fig.~\ref{fig:toy_NNactivations}.  In the top two examples, the network finds the signal region and correctly estimates the magnitude of the likelihood ratio. In both these cases, the network also overtrains on a (real) fluctuation in the training data, despite the validation procedure. Such regions will tend to decrease the effectiveness of the classifier, since a given cut threshold will admit more background in the test data. In the bottom left example of Fig.~\ref{fig:toy_NNactivations}, the network finds a function approximately monotonic to $h(x,y)$ but with different normalization -- while the cost function would have preferred to optimize this network to reach $h(x,y)$, the validation procedure cut off the optimization when the correct shape to isolate the signal region had been found. Due to the nature of the cuts, there is no performance loss for this network, since crucially it has found the correct shape near the signal region. The last network fails to converge to the signal region, and instead focuses its attention on the fluctuation in the training data.  The variation in the network performance illustrates the importance of training multiple classifiers and using schemes to mitigate the impact of statistical fluctuations in the training dataset.


\begin{figure}[h!]
\centering
\includegraphics[width=0.6\textwidth]{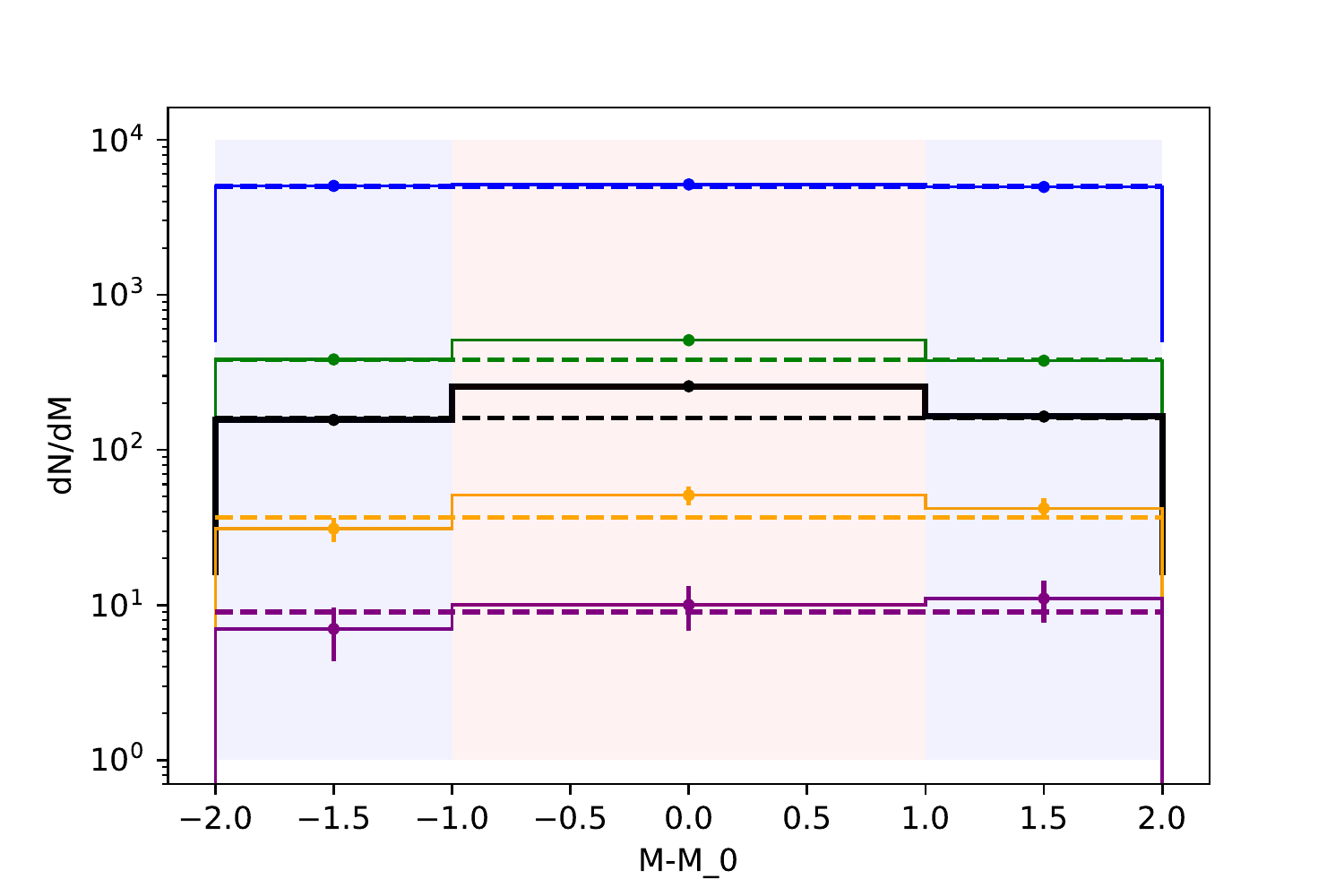}
\caption{The mass distribution after various threshold on the neural network classifier. The flat background fit from the sideband regions are the dashed lines, and the statistical uncertainty for each bin is shown by the error bars. The top histogram is the model before any threshold in the $(x,y)$ plane, and from top to bottom respectively the histogram is given for efficiency thresholds of $10\%,5\%,1\%,0.2\%$. The significance is $\mathcal{S} = 3\sigma, 9.4\sigma,  10.8\sigma,$ and  $3.4\sigma$ for respectively no threshold, $10\%$, $5\%$, and $1\%$. The $0.2\%$ threshold reduces the signal to no statistical significance. \label{fig:toy_cuts}}
\end{figure}

Figure~\ref{fig:toy_cuts} shows the mass distribution in the three bins after applying successfully tighter threshold on the neural network output.  Since $Y$ is not a truth bit, the data are reduced in both the signal region and the mass sidebands.  For each threshold, the background expectation $\hat{n}_b$ assuming a uniform distribution is estimated by fitting a straight line to the mass sidebands.  Then, the significance is estimated from the number of observed events in the signal region, $n_o$, via $\mathcal{S}\approx (n_o-\hat{n}_b)/\sqrt{\hat{n}_b}$.  Of the threshold presented, the maximum significance corresponds to the 5\% efficiency with $\mathcal{S}\approx 10.8\sigma$.  Even though the ideal significance is $15\sigma$, for the particular pseudodataset shown in Fig.~\ref{fig:toy_cuts}, the ideal classifier significance is $13.9\sigma$.

\begin{figure}[h!]
\centering
\includegraphics[width=0.45\textwidth]{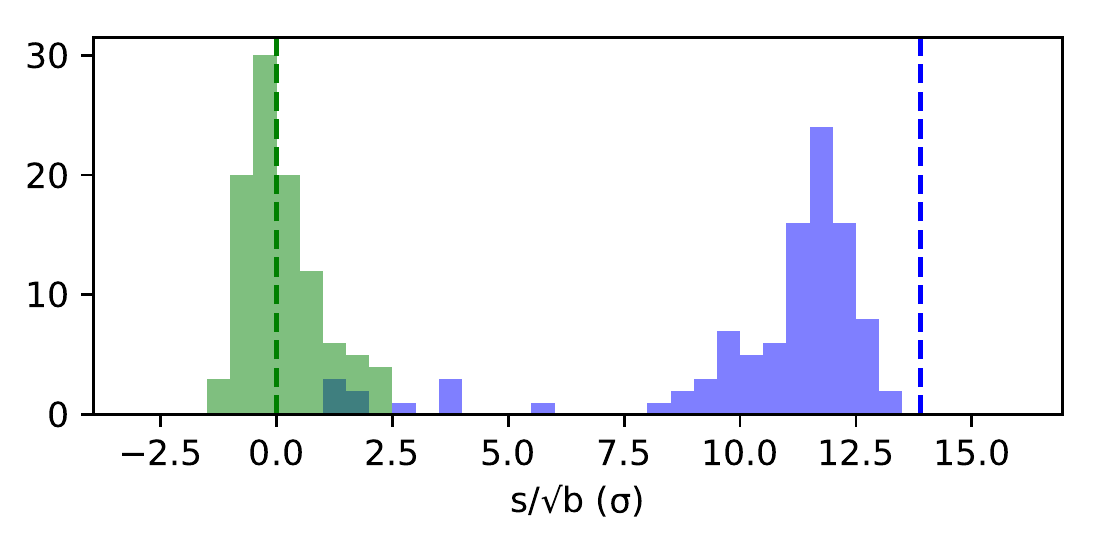}
\includegraphics[width=0.45\textwidth]{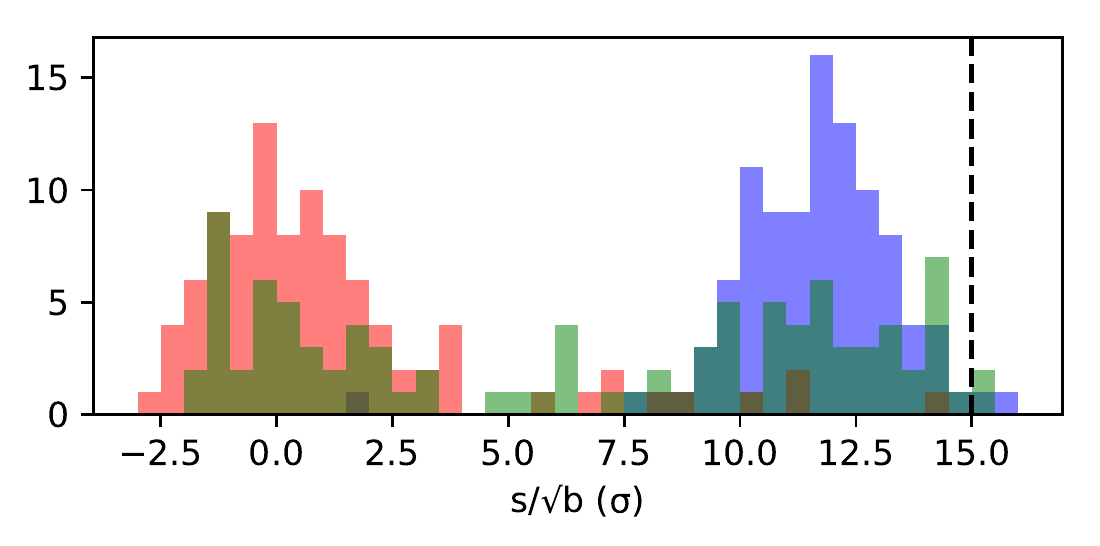}
\includegraphics[width=0.45\textwidth]{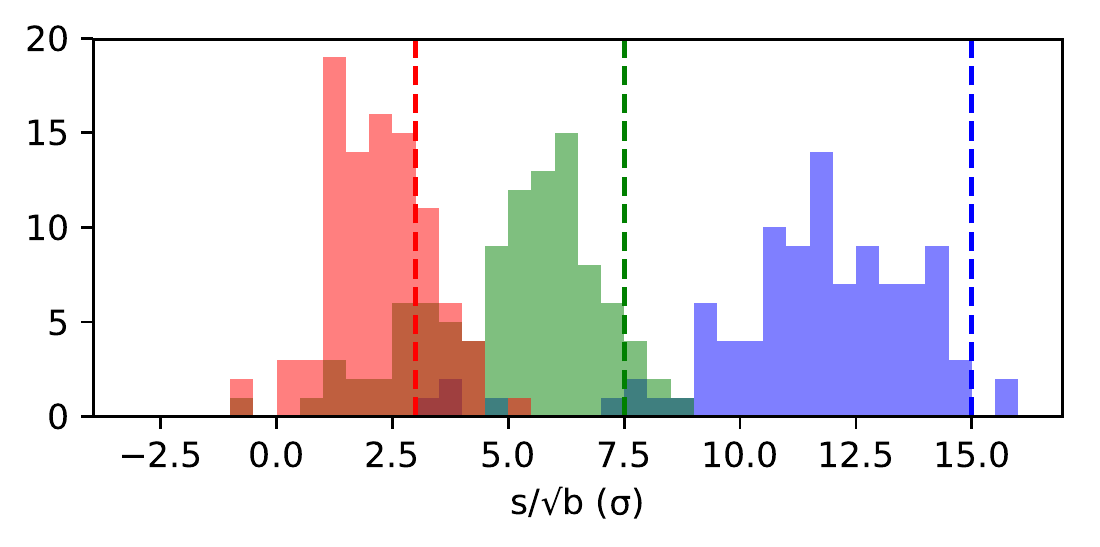}
\caption{\textbf{Top-left:} histogram of significance at a test theshold of $6\%$ for 100 NN trained on the example toy model data (blue), and 100 NN trained on a control dataset with no signal present (green). The dashed green line gives the expected significance of $13.9\sigma$ for the example dataset with ideal cuts.  
\textbf{Top-right:} histogram of significance for ensembles with expected signal strength of $15\sigma$ with ideal cuts. The blue is $(N_b = 10000, N_s = 300, w_s = 0.2)$ at a test threshold of $6\%$, the green is $(N_B=10000, N_s = 150, w_s = 0.1)$ at a test threshold of $1.5\%$, and the red is $(N_B=10000, N_s = 75, w_s = 0.05)$ at a test threshold of $0.4\%$. For each ensemble, 100 independent instances of the dataset are generated and one NN is trained on each dataset.
\textbf{Bottom:} histogram of significance for ensembles with $(N_b = 10000, N_s = 300)$ and varying $w_s$. Blue is $w_s = 0.2$ at a test threshold of $6\%$, green is $w_s=0.4$ at a test threshold of $24\%$, and red is $w_s=1.0$ (for which the background and signal distribution in $(x,y)$ are identical) at a test threshold of $50\%$. For each ensemble, 100 independent instances of the dataset are generated and one NN is trained on each dataset. The dashed line gives the expected significance for each ensemble.}
\label{fig:toy_ensembles}
\end{figure}

We can study the behavior of our NN classifiers by looking at the significance generated by ensembles of models trained on signals of different strength, as shown in Fig.~\ref{fig:toy_ensembles}. The top histogram shows the significance for an ensemble of models trained on the example signal (blue) and on a control dataset with no signal (green). The control ensemble appears to be normally distributed around $s/\sqrt{b}=0$, while the example signal ensemble is approximately normally distributed around $12\sigma$ (compared to $13.9\sigma$ for the ideal cut), along with a small $O(5\%)$ population of networks that fail to find the signal. The middle histogram shows the effect of decreasing the size of the signal region $w_s$ while modifying $N_s$ to maintain an expected significance of $15\sigma$ with ideal cuts. When $w_s$ is decreased, the training procedure appears to have a harder time picking up the signal, possibly due to our choice of an operating point of $2\%$ false-positive rate for the sideband validation. For $w_s=0.1$ (green), about $50\%$ of the networks effectively find the signal. For $w_s =0.05$ (red), only about $5\%$ find the signal. The bottom plot shows the effect of increasing $w_s$ while keeping $N_s$ fixed, so that the strength of the signal decreases. When the size of the signal region is doubled to $w_s=0.4$ (green), giving a expected signifance of $7.5\sigma$, the network performs similarly to the $w_s=0.2$ example (blue). When the signal distribution is identical to the background distribution ($w_s=1.0$, red), there is on average a small decrease in performance compared to simply not using a classifier.

\section{Full Method}
\label{sec:fullmethod}

The previous sections uses key elements of the full extended bump hunt but do not include all components, including the full background estimation and statistical analysis.  This section gives a concrete prescription for applying the CWoLa hunting method in practice, which will be used in an explicit example in Sec.~\ref{sec:physicsexample}.  The setup is as in the previous sections: there is feature $m_\text{res}$ where the signal is expected to be resonant and then a set of other features $Y$ that are uncorrelated with $m_\text{res}$, but potentially useful for distinguishing signal from background.  It is important to state that while a detailed model of $Y$ is not required to perform the CWoLa hunting procedure that is described in the rest of the section, a limited model of $Y$ is required to ensure the correlations with $m_\text{res}$ are minimal.  Such a model could come from simulation, from theory, or directly from a sufficiently signal-devoid data sample.

While in the presence of signal, the CWoLa hunting method would ideally learn systematic correlations between $m_\text{res}$ and $Y$, instead, it may focus on statistical fluctuations in the background distributions. A naive application of CWoLa directly on the data may produce bumps in $m_\text{res}$ by seeking local statistical excesses in the background distribution. This corresponds to a large look-elsewhere effect over the space of observables $Y$ -- the classifier may search this entire space and find the selection with the largest statistical fluctuation. In Sec.~\ref{sec:example}, we took the approach of splitting the dataset into training, validation and test samples which eliminates this affect, since the statistical fluctuations in the three samples will be uncorrelated. However, applying this approach in practice would reduce the effective luminosity available for the search and thus degrade sensitivity. We therefore apply a cross-validation technique which allows all data to be used for testing while ensuring that event subsamples are never selected using a classifier that was trained on them. We split the events randomly, bin-by-bin, into five event samples of equal size. The first sample is set aside, and the first classifier is trained on the signal- and sideband-region events of the remaining four samples. This classifier may learn the statistical fluctuations in these event samples, but those will be uncorrelated with the fluctuations of the first sample. Applying the classifier to the set-aside event sample will then correspond to only one statistical test, eliminating the look elsewhere effect. By repeating this procedure five times -- each time setting aside one $k$-fold for testing and four for training and validation, all the data can be used for the bump hunt by adding up the selected events from each $k$-fold.

\begin{algorithm}[t]
\label{algo:crossval}
Split dataset into 5 subsets stratified by $m_\text{res}$ binning

 \For{$\mathrm{subset}_i$ in subsets}{
 Set aside $\mathrm{subset}_i$ as test data
 
  \For{$\mathrm{subset}_j$ in subsets, $j \ne i$}{
    Validation data sideband = merge sideband bins of $\mathrm{subset}_j$

	Validation data signal-region = merge signal-region bins of $\mathrm{subset}_j$
	
	Training data sideband = merge sideband bins of remaining subsets

	Training data signal-region = merge signal-region bins of remaining subsets
	
	Assign signal-region data with label 1
	
	Assign sideband data with label 0
	
	Train twenty classifiers on training data, each with different random initialization
	
	$\mathrm{model}_{i, j}$ = best of the twenty models, as measured by performance on validation data
  }
  $\mathrm{model}_{i}$ = $\sum_j \mathrm{model}_{i,j} / 4$
  
  Select $Y$\% most signal-like data points of $\mathrm{subset}_i$, as determined by $\mathrm{model}_{i}$.  The threshold on the neural network to achieve $Y\%$ is determined using all other bins with large numbers of events and so the uncertainty on the value is negligible.
 }
 
Merge selected events from each subset into new $m_\text{res}$ histogram

Fit smooth background distribution to $m_\text{res}$ distribution with the signal region masked

Evaluate $p$-value of signal region excess using fitted background distribution interpolated into the signal region.

 \caption{Nested cross-validation training and event selection procedure.  See Appendix~\ref{app:stats} for further details on the last two points.}
\end{algorithm}

\begin{figure}[t]
\centering
\includegraphics[width=0.95\textwidth]{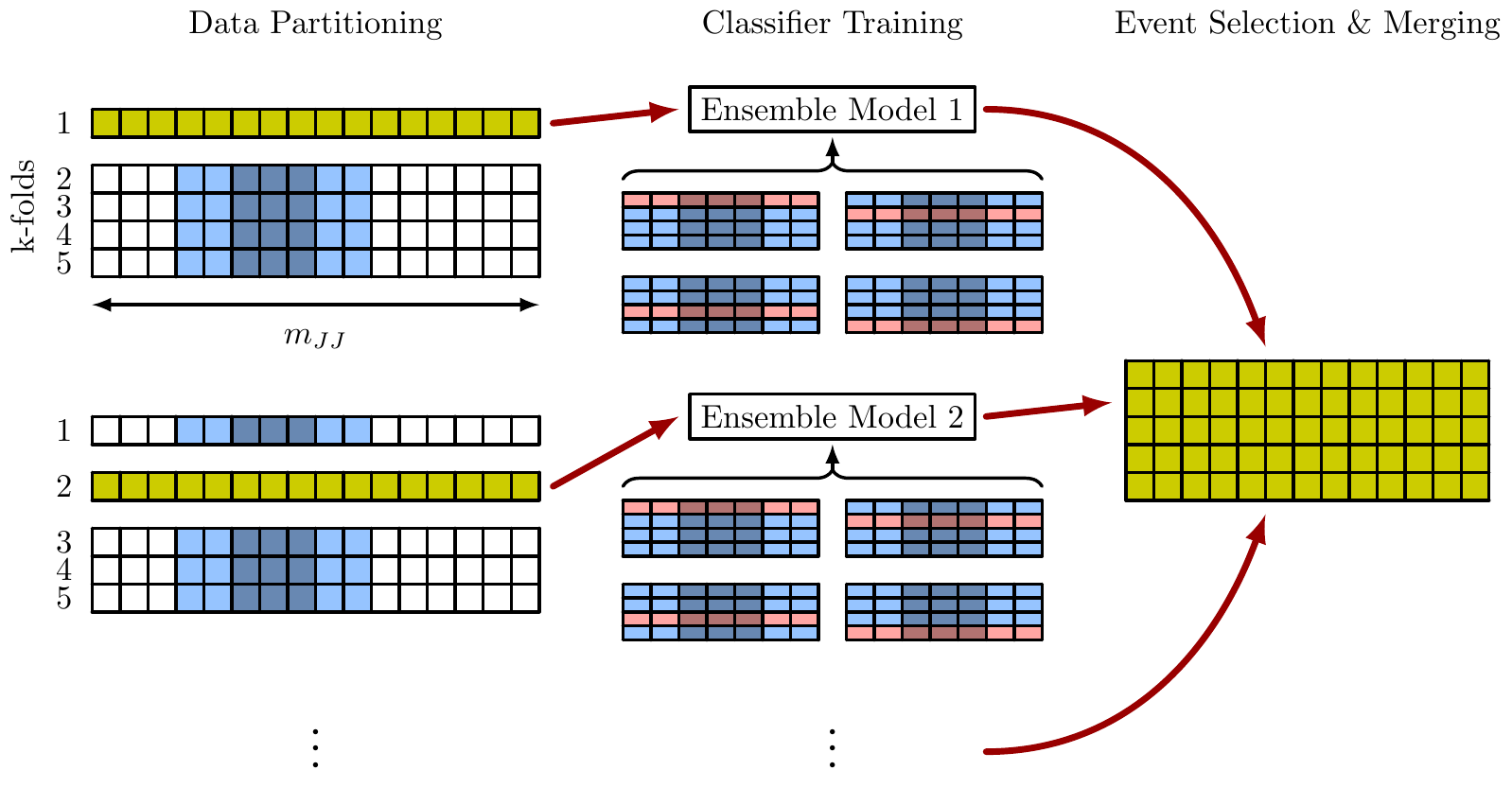}
\caption{Illustration of the nested cross-validation procedure. \textbf{Left:} the dataset is randomly partitioned bin-by-bin into five groups. \textbf{Center:} for each group $i\in\{1,2,3,4,5\}$ (the test set), an ensemble classifier is trained on the remaining groups $j\neq i$. There are four ways to split the four remaining groups into three for training and one for validation.  For each of these four ways, many classifiers are trained and the one with best validation performance is selected. The ensemble classifier is then formed by the average of the four selected classifiers (one for each way to assign the training/validation split). \textbf{Right:} Data are selected from each test group using a threshold cut from their corresponding ensemble classifier. The selected events are then merged into a single $m_\text{res}$ histogram.}
\label{fig:crossval}
\end{figure}

The algorithm we used for this procedure is summarized in Algorithm~\ref{algo:crossval}, and illustrated in Fig.~\ref{fig:crossval}. For each set-aside test set, we perform four rounds of training and validation using the four remaining data subsets. In each round, we set aside one of the remaining subsets as validation data, and the final three are used for training data. Only data falling in the signal and sideband regions are used for training and validation. The training and validation data are labelled as 0 or 1 if they fall in the sideband or signal regions, respectively. For each round, we train twenty NNs on the same training and validation data, using a different initialization each time. Each classifier is validated according to its performance as measured on validation data. Our performance metric $\epsilon_\mathrm{val}$ is the true positive rate for correctly classifying a signal-region event as such, evaluated at a threshold with given false positive rate $s\%$ for incorrectly classifying a sideband region event as a signal region event. If a signal is present in the signal region and the classifier is able to find it, then it should be that $\epsilon_\mathrm{val} > s\%$. On the other hand, if no signal is present then $\epsilon_\mathrm{val} \simeq s\%$ is expected. Since we will be typically considering $\mathcal{O}(1\%)$-level signals we consider $s\% \sim 1\%$ in our test, and set $s\% = 0.5\%$ to generate our final results. For each of the twenty models, we end training if its performance has not improved in 300 epochs, and revert the model to a checkpoint saved at peak performance. We select the best of these twenty models, and discard the others. At the end of four rounds, the four selected models are are averaged to form an ensemble model which is expected to be more robust than any individual model. The ensemble model is used to classify events in the test set, by selecting the $r\%$ most signal-like events. This procedure is repeated for all five choices of test set, and the selected events from each are combined into a signal histogram in $m_\text{res}$. The presence of an identifiable signal will be indicated by a bump in the signal region, for which standard bump-hunting techniques can be used to perform a hypothesis test. The use of averaged ensemble models is important to reduce any performance degradation due to overfitting. Since each of the four models used to make each ensemble model has been trained on different training sets and with different random weight initialization, they will tend to overfit to different events. The models will therefore disagree in regions where overfitting has occurred, but will tend to agree in any region where a consistent excess is found.

Further technical details about the statistical methods can be found in Appendix~\ref{app:stats}.  Asymptotic formulae can be used to determine the local $p$-value of an excess, but such formulae must be validated using more computationally expensive methods for each application of CWoLa hunting, as is demonstrated in the appendix.

\subsection{Interpreting the Results}

The main result following the application of the method from Sec.~\ref{sec:fullmethod} is the local $p$-value.  To determine the compatibility of the entire mass range with the no-resonance hypothesis, it is desirable to be able to compute a global $p$-value.  In the result presented here, the mass bins were fixed ahead of time and were also non-overlapping.  Therefore, it is relatively simple to estimate a global $p$-value using e.g. a Bonferroni correction.  However, this is not ideal (over-conservative) when the mass bin width is scanned as part of the procedure.  It is still possible to determine a global $p$-value, in the same spirit as the full bumphunter statistic~\cite{Choudalakis:2011qn}.  This would require a significant computational overhead as a large number of neural networks would need to be trained for each of many pseudo-experiments.  An additional trials factor would be associated with scanning the threshold fraction on the neural network output.  In the simplest approach, a small number of well-separated working points would be chosen, such as 10\%, 1\%, and 0.1\%.  These should be sufficiently different that the three local $p$-values could be treated as independent.  However, a finer scan would require a proper assessment of the global $p$-value using pseudo-experiments.  It may be possible to significantly reduce the computational cost by estimating the correlation between mass windows and threshold fractions in order to properly account for the look-elsewhere-effect~\cite{Gross:2010qma,VITELLS2011230}.  


One final remark is about how one would use CWoLa hunting to set limits.  In the form described above, the CWoLa hunting approach is designed to find new signals in data without any model assumptions.  However, it is also possible to recast the lack of an excess as setting limits on particular BSM models.  Given a simulated sample for a particular model, it would be possible to set limits on this model by mixing the simulation with the data and training a series of classifiers as above and running toy experiments, re-estimating the background each time.  This is similar to the usual bump hunt, except that there is more computational overhead because the background distribution is determined in part by the neural networks, and the distribution in expected signal efficiencies cannot be determined except by these toy experiments\footnote{This complicates the legacy utility of the results, but it would be possible to tweak procedures like those advocated by RECAST~\cite{Cranmer:2010hk} in which neutral networks would be automatically trained for a new signal model.}.  In the absence of an excess, it is also possible to directly recast the results by taking the classifier trained on data with no significant signal.  However, without a real excess, the classifier will have nothing to learn.  Such a classifier will likely not be useful for any particular signal model.  Therefore, while it is technically possible to do a standard re-interpretation of the results, the most powerful limit setting requires access to the data to retrain the neural networks for an injected signal.

\section{Physical Example}
\label{sec:physicsexample}

This section uses a dijet resonance search at the LHC to show the potential of CWoLa hunting in a realistic setting.  As discussed in Sec.~\ref{sec:intro}, both ATLAS and CMS have a broad program targeting resonance decays into a variety of SM particles.  Due to significance advances in jet substructure-based tagging~\cite{Larkoski:2017jix}, searches involving hadronic decays of the SM particles can be just as if not more powerful than their leptonic counterparts.  The usual strategy for these searches is to develop dedicated single-jet classifiers, including\footnote{These are the latest $\sqrt{s}=13$ TeV results - see references within to find the complete history.} $W/Z$-~\cite{CMS-DP-2015-043,ATLAS-CONF-2017-064}, $H$-~\cite{CMS-DP-2015-038,ATLAS-CONF-2016-039}, top-~\cite{CMS-DP-2015-043,CMS-PAS-JME-15-002,ATLAS-CONF-2017-064}, $b$-~\cite{CMS-DP-2017-012,ATL-PHYS-PUB-2017-013}, and quark-jet taggers~\cite{CMS-DP-2016-070,ATL-PHYS-PUB-2017-009}.  Simulated events with per-instance labels are used for training and then these classifiers are deployed in data.  However, the best classifier in data may not be the best classifier in simulation.  This problem is alleviated when learning directly from data.  

Learning directly from data has another advantage - the decay products of a new heavy resonance may themselves be beyond the SM.  If the massive resonance decays into new light states such as BSM Higgs bosons or dark sector particles that decay hadronically, then no dedicated SM tagger will be optimal~\cite{Aguilar-Saavedra:2018xpl,Aguilar-Saavedra:2017zuc}.  A tagger trained to directly find non-generic-jet structure could find these new intermediate particles and thus also find the heavy resonance. This was the approach taken in~\cite{Aguilar-Saavedra:2017rzt}, but that method is fully supervised and so suffers the usual theory prior bias and potential sources of mismodelling.  Here we will illustrate how the CWoLa hunting approach could be used instead to find such a signal.  The next section (Sec.~\ref{sec:simulation}) describes the benchmark model in more detail, as well as the simulation details for both signal and background.

\subsection{Signal and background simulation}
\label{sec:simulation}

For a benchmark signal, we consider the process $p p \to W' \to W X, X \to W W$, where $W'$ and $X$ are a new vector and scalar particle respectively. This process is predicted, for example, in the warped extra dimensional construction of \cite{Agashe:2016rle, Agashe:2017wss, boosted_diboson}. The typical opening angle between the two $W$ bosons resulting from the $X$ decay is given by $\Delta R(W,W) \simeq 4 \, m_{X} / m_{W'}$ for $2 m_W \ll m_X \ll m_{W'}$, and so the $X$ particle will give rise to a single large-radius jet in the hadronic channel when $m_X \lesssim m_{W'}/4$. Taking the mass choices $m_{W'} = 3 \; \text{TeV}$ and $m_X = 400 \; \text{GeV}$, the signal in the fully hadronic channel is a pair of large-radius jets $J$ with $m_{JJ} \simeq 3 \; \text{TeV}$, one of which has a jet mass $m_J \simeq 80 \; \text{GeV}$ and a two-pronged substructure, and the other has mass $m_J \simeq 400 \; \text{GeV}$ with a four-prong substructure which often is arranged as a pair of two-pronged subjets.

Events are generated with Madgraph5\_aMC@NLO~\cite{Alwall:2014hca} v2.5.5 to generate $10^4$ signal events, using a model file implementing the tensor couplings of~\cite{Agashe:2017wss} and selecting only the fully hadronic decays of the three $W$ bosons. The events are showered using Pythia 8.226~\cite{Sjostrand:2007gs}, and are passed through the fast detector simulator Delphes 3.4.1~\cite{deFavereau:2013fsa}. Jets are clustered from energy-flow tracks and towers using the FastJet~\cite{Cacciari:2011ma} implementation of the anti-$k_t$ algorithm \cite{Cacciari:2008gp} with radius parameter $\Delta R = 1.2$. We require events to have at least two ungroomed large-radius jets with $p_T > 400 \; \text{GeV}$ and $|\eta| < 2.5$. The selected jets are groomed using the soft drop algorithm~\cite{Larkoski:2014wba} in grooming mode, with $\beta = 0.5$ and $z_\text{cut} = 0.02$. The two hardest groomed jets are selected as a dijet candidate, and a suite of substructure variables are recorded for these two jets. With the same simulation setup, $4.45 \times 10^6$ Quantum Chromodynamic (QCD) dijet events are generated with parton level cuts $p_{T, \, j} > 300 \; \text{GeV}$, $|\eta_j| < 2.5$, $m_{jj} > 1400 \; \text{GeV}$.  

In order to study the behaviour of the CWoLa hunting procedure both in the presence and absence of a signal, we produce samples both with and without an injected signal. The events are binned uniformly in $\log(m_{JJ})$, with 15 bins in the range $2001 \; \text{GeV} < m_{JJ}  < 4350 \; \text{GeV}$.

\subsection{Training a Classifier}
In order to test for a signal with mass hypothesis $m_{JJ} \simeq m_{\text{res}}$, we construct a `signal region' consisting of all the events in the three bins centered around $m_{\text{res}}$. We also construct a low- and a high-mass sideband consisting of the events in the two bins below and above the signal region, respectively. The mass hypothesis will be scanned over the range $2278 \; \text{GeV} \leq m_\text{res} \leq 3823 \; \text{GeV}$, to avoid the first and last bins that can not have a reliable background fit without constraints on both sides of the signal region.  The signal region width is motivated by the width of the $m_{JJ}$ peak for the benchmark signal process described earlier. Because all particles in the process are very narrow, this width corresponds to the resolution allowed by the jet reconstruction and detector smearing effects and will be relevant for other narrow signal processes also. For processes giving rise to wider bumps, the width of the signal hypothesis could be scanned over just as we scan over the mass hypothesis. We will then train a classifier to distinguish the events in the signal region from those in the sideband on the basis of their substructure. The objective in constructing the training framework is that the classifier should be very poor (equal efficiency in signal region and sideband for any threshold) in the case that no signal is present in the signal region, but if a signal is present with unusual jet substructure then the classifier should be able to locate the signal and provide discrimination power between signal and SM dijet events.

The background is estimated by fitting the regions outside of the signal region to a smoothly falling distribution.  In practice, this requires that the auxiliary information $Y$ is nearly independent of $m_{JJ}$; otherwise, the distribution could be sculpted.  To illustrate the problem, consider a classifier trained to distinguish the sideband and signal regions using the observables $m_J$ and the N-subjettiness variable $\tau_1^{(2)}$~\cite{Thaler:2011gf}.  The ratio $m_J /  \sqrt{\tau_1^{(2)}}$ is approximately the jet $p_\text{T}$, which is highly correlated with $m_{JJ}$ for the background.    While it is often possible to find ways to decorrelate substructure observables~\cite{Dolen:2016kst,Shimmin:2017mfk,Aguilar-Saavedra:2017rzt,Moult:2017okx}, we take a simpler approach and instead select a basis of substructure variables which have no strong correlations with $m_{JJ}$. We will use the following set of 12 observables which does not provide learnable correlations with $m_{JJ}$ sufficient to create signal-like bumps in our simulated background dijet event samples, as we shall demonstrate later in this section
\begin{equation}
\text{For each jet:} ~~~~~Y_i =  \left(m_J, ~\sqrt{\tau_1^{(2)}} / \tau_1^{(1)},~ \tau_{21}, ~\tau_{32},~ \tau_{43},~ n_\text{trk}\right),
\end{equation}
where $\tau_{MN} = \tau_M^{(1)} / \tau_N^{(1)}$. The full training uses $Y=(Y_1,Y_2)$.  All ratios of N-subjettiness variables are chosen to be invariant under longitudinal boosts, so that the classifier cannot learn $p_T$ from $m_J$ and the other observables. The two jets are ordered by jet mass, so that the first six observables $Y_1$ correspond to the heavier jet while the last six $Y_2$ correspond to the lighter jet. We find that while the bulk of the $m_J$ distribution in our simulated background dijet samples do not vary strongly over the sampled range of $m_{JJ}$, the high mass tails of the heavy and light jet mass distributions are sensitive to $m_{JJ}$. In lieu of a sophisticated decorrelation procedure, we simply reject outlier events which have $m_{J,\,A} > 500 \; \text{GeV}$ and $m_{J,\,B} > 300 \; \text{GeV}$, where the subscripts $A$ and $B$ refer to the heavier and lighter jet respectively.

In our study, the classifiers used are dense neural networks built and trained using Keras with a TensorFlow backend. We use four hidden layers consisting of a first layer of 64 nodes with a leaky Rectified Linear Unit (ReLU) activation (using an inactive gradient of 0.1), and second through fourth layers of 32, 16, 4 nodes respectively with Exponential Linear Unit (ELU) activation~\cite{clevert2015fast}. The output node has a sigmoid activation. The first three hidden layers are regulated with dropout layers with 20\% dropout rate~\cite{JMLR:v15:srivastava14a}. The neural networks are trained to minimize binary cross-entropy loss using the Adam optimizer with learning rate of 0.001, batch size of 20000, first and second moment decay rates of 0.8 and 0.99, respectively, and learning rate decay of $5\times10^{-4}$. The training data is reweighted such that the low sideband has equal total weight to the high sideband, the signal region has the same total weight as the sum of the sidebands, and the sum of all events weights in the training data is equal to the total number of training events. This ensures that the NN output will be peaked around 0.5 in the absence of any signal, and ensures that low and high sideband regions contribute equally to the training in spite of their disparity in event rates.

\subsection{Results}

We use a sample of 553388 QCD dijet events with dijet invariant mass $m_{JJ} > 2001 \; \mathrm{GeV}$, corresponding to a luminosity of $4.4 \; \mathrm{fb}$. We consider two cases: first, a background-only sample; and second, a sample in which a signal has been injected with $m_{JJ} \simeq 3000 \; \text{GeV}$, with 877 events in the range $m_{JJ} > 2001 \; \mathrm{GeV}$. In the signal region $2730 \; \mathrm{GeV} < m_{JJ} < 3189 \; \mathrm{GeV}$, consisting of the three bins centered around $3000 \; \text{GeV}$, there are 81341 background events and 522 signal events, corresponding to $S/B = 6.4\times10^{-3}$ and $S/\sqrt{B} = 1.8$. Labelling the bins 1 to 15, we perform the procedure outlined previously to search for signals in the background-only and background-plus-signal datasets in signal regions defined around bins 4 - 12. This leaves room to define a signal region three bins wide, surrounded by a low and high sideband each two bins wide.

\begin{figure}[!t]%
\centering
\includegraphics[width=\textwidth]{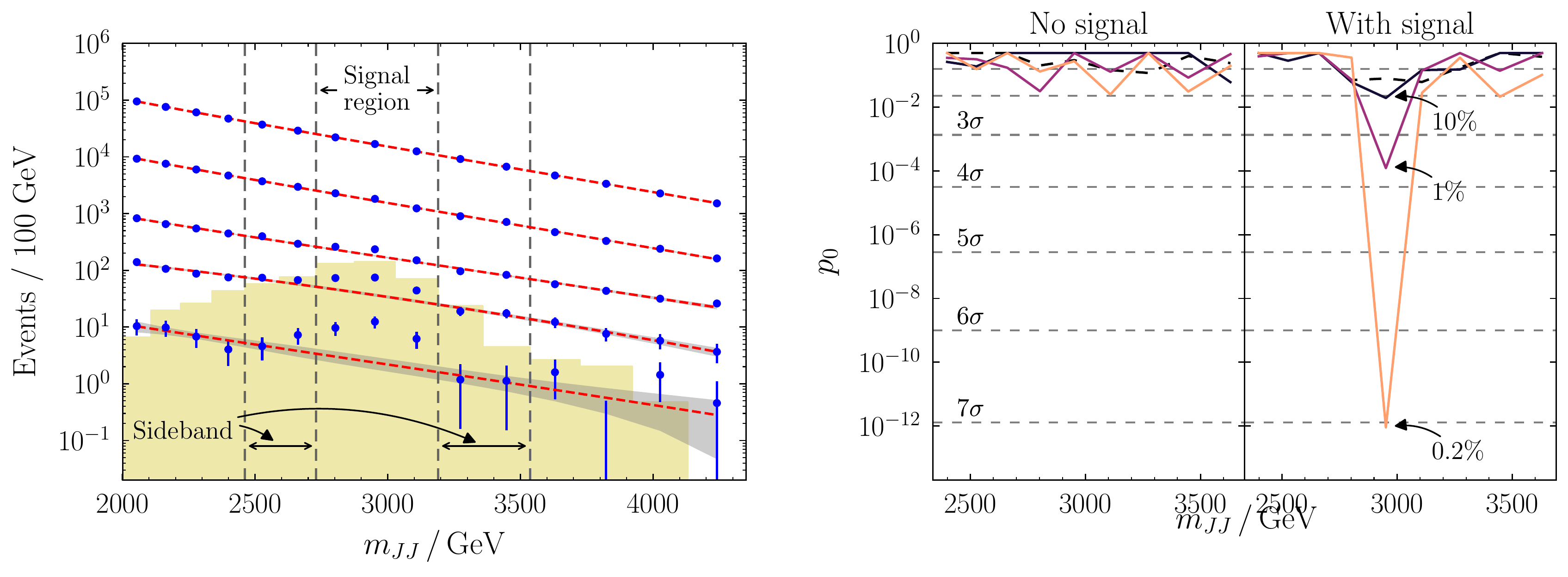}
\caption{\textbf{Left:} $m_{JJ}$ distribution of dijet events (including injected signal, indicated by the filled histogram) before and after applying jet substructure cuts using the NN classifier output for the $m_{JJ} \simeq 3 \; \text{TeV}$ mass hypothesis. The dashed red lines indicate the fit to the data points outside of the signal region, with the gray bands representing the fit uncertainties. The top dataset is the raw dijet distribution with no cut applied, while the subsequent datasets have cuts applied at thresholds with efficiency of $10^{-1}$, $10^{-2}$, $2\times10^{-3}$, and $2\times10^{-4}$. \textbf{Right:} Local $p_0$-values for a range of signal mass hypotheses in the case that no signal has been injected (left), and in the case that a $3 \; \text{TeV}$ resonance signal has been injected (right). The dashed lines correspond to the case where no substructure cut is applied, and the various solid lines correspond to cuts on the classifier output with efficiencies of $10^{-1}$, $10^{-2}$, and $2\times10^{-3}$.}%
\label{fig:pvalues}%
\end{figure}

In Fig.~\ref{fig:pvalues} left, we plot the the background-plus-signal dataset which survive cuts at varying thresholds using the output of the classifier trained on the signal bin centered around $3 \; \mathrm{TeV}$. The topmost distribution corresponds to the inclusive dijet mass distribution, while the subsequent datasets have thresholds applied on the neural network with overall efficiencies of 10\%, 2\%, and 0.5\%, respectively. A clear bump develops at the stronger thresholds, indicating the presence of a $3 \; \mathrm{TeV}$ resonance. The automated procedure used to determine the significance is explained in detail in Appendix~\ref{app:stats}. In brief, we estimate the background in the signal region by performing a fit of a smooth three-parameter function to the event rates in all the bins besides those in the signal region.  We perform a simple counting experiment in the signal region, using the profile likelihood ratio as the test statistic, with the background fit parameters treated as nuisance, with pre-profile uncertainties taken from the background fit itself.  The significance is estimated using asymptotic formulae describing the properties of the profile likelihood ratio statistic~\cite{Cowan:2010js}.   Figure~\ref{fig:pvalues} shows the signal significance for each signal mass hypothesis, in the case that no signal is present (left), and in the case that the signal is present (right). We see that when no signal is present, no significant bump is created by our procedure. When a signal is present with $m_{\text{res}} = 3 \; \mathrm{TeV}$, there is a significant bump which forms at this signal hypothesis, reaching $7\sigma$ at 0.2\% efficiency. In Appendix~\ref{app:scan_plots}, we show the $m_{JJ}$ distributions for each scan point used for the calculation of these $p$-values.

The fact that there is no significant bump in the left plot of Fig.~\ref{fig:pvalues} is an important method closure test.  When deploying the CWoLa hunting approach in practice, we advocate to test the method in simulation in order to validate that there are no bump-catalyzing correlations in the selected classification features.  A residual concern may be that there are correlations in the data which are not present in simulation.  Residual correlations may come in two forms: process and kinematic.  Process correlations occur when $Y$ depends on the production channel (e.g. $pp\rightarrow qq$ or $pp\rightarrow gg$) and $m_{JJ}$ also depends on the production channel; kinematic correlations are the case when $m_{JJ}$ is correlated with $Y$ given the process.  Residual process correlations do not cause bumps because the $m_{JJ}$ distribution of each process type (aside from signal) is smoothly falling.  Thus, even if the classifier can exactly pick out one process, no bumps will be artificially sculpted.  Residual kinematic correlations could cause artificially bumps in the $m_{JJ}$ distribution.  Physically, kinematic correlations occur because $Y_i$ is correlated with $p_\text{T,i}$.  One way to show in data that residual kinematic correlations are negligible is to use a \textit{mixed sample} in which pairs of jets from different events are combined.  As long as the potential signal fraction is small, this mixed sample will have no resonance peak.  While the features chosen in this section were designed to be uncorrelated with $m_{JJ}$ and not sculpt bumps, it may be possible to utilize correlated features in a modified CWoLa hunting procedure that includes systematic uncertainties for strong residual correlations.  We leave studies of this possibility to future work.

\begin{figure}[!t]
\centering
\includegraphics[width=\textwidth]{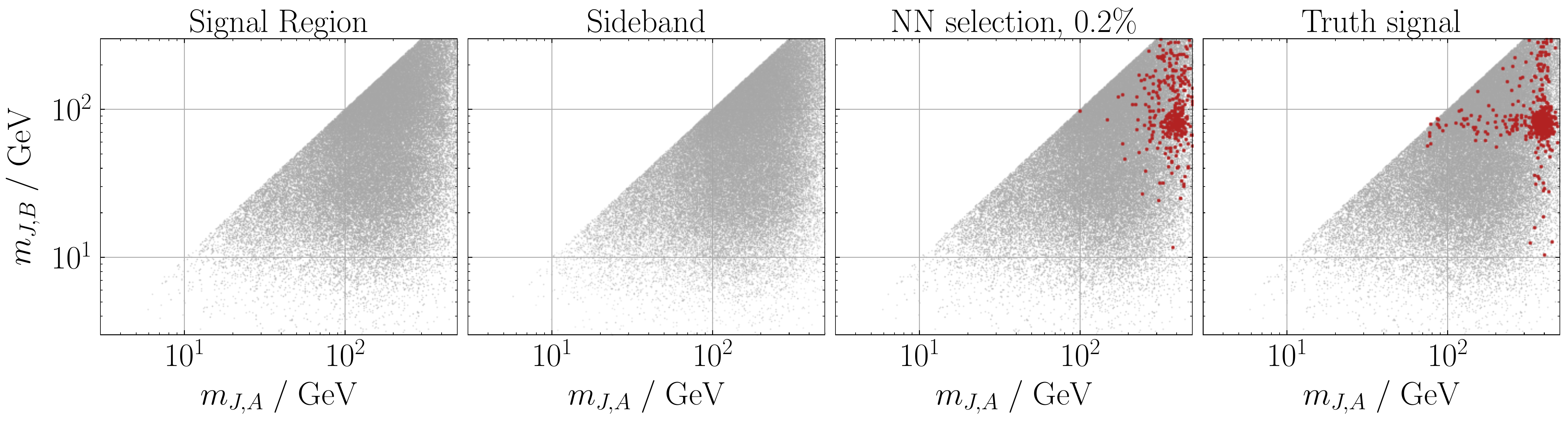}
\caption{Events projected onto the 2D plane of the two jet masses. The classifiers are trained to discriminate events in the signal region (left plot) from those in the sideband (second plot). The third plot shows in red the 0.2\% most signal-like events determined by the classifier trained in this way. The rightmost plot shows in red the truth-level signal events.}
\label{fig:scattermJJ}
\end{figure}

We can investigate what the classifier has learnt by looking at the properties of events which have been classified as signal-like. In the first (second) plot of Fig.~\ref{fig:scattermJJ}, events in the signal (sideband) region have been plotted on the plane of the jet masses of the heavier jet ($m_{J,\,A}$) and the lighter jet ($m_{J,\,B}$). After being trained to discriminate the events of the signal region from those of the sideband, the 0.2\% most signal-like events as determined by the classifier are plotted in red in the third plot of Fig.~\ref{fig:scattermJJ}, overlaid on top of the remaining events in gray. The classifier has selected a population of events with $m_{J\,A} \simeq 400 \; \mathrm{GeV}$ and $m_{J\,B} \simeq 80 \; \mathrm{GeV}$, consistent with the injected signal. The final plot of the figure shows in red the truth-level signal events, overlaid on top of the truth-level background in grey.

\begin{figure}[p]
\centering
\includegraphics[width=\textwidth]{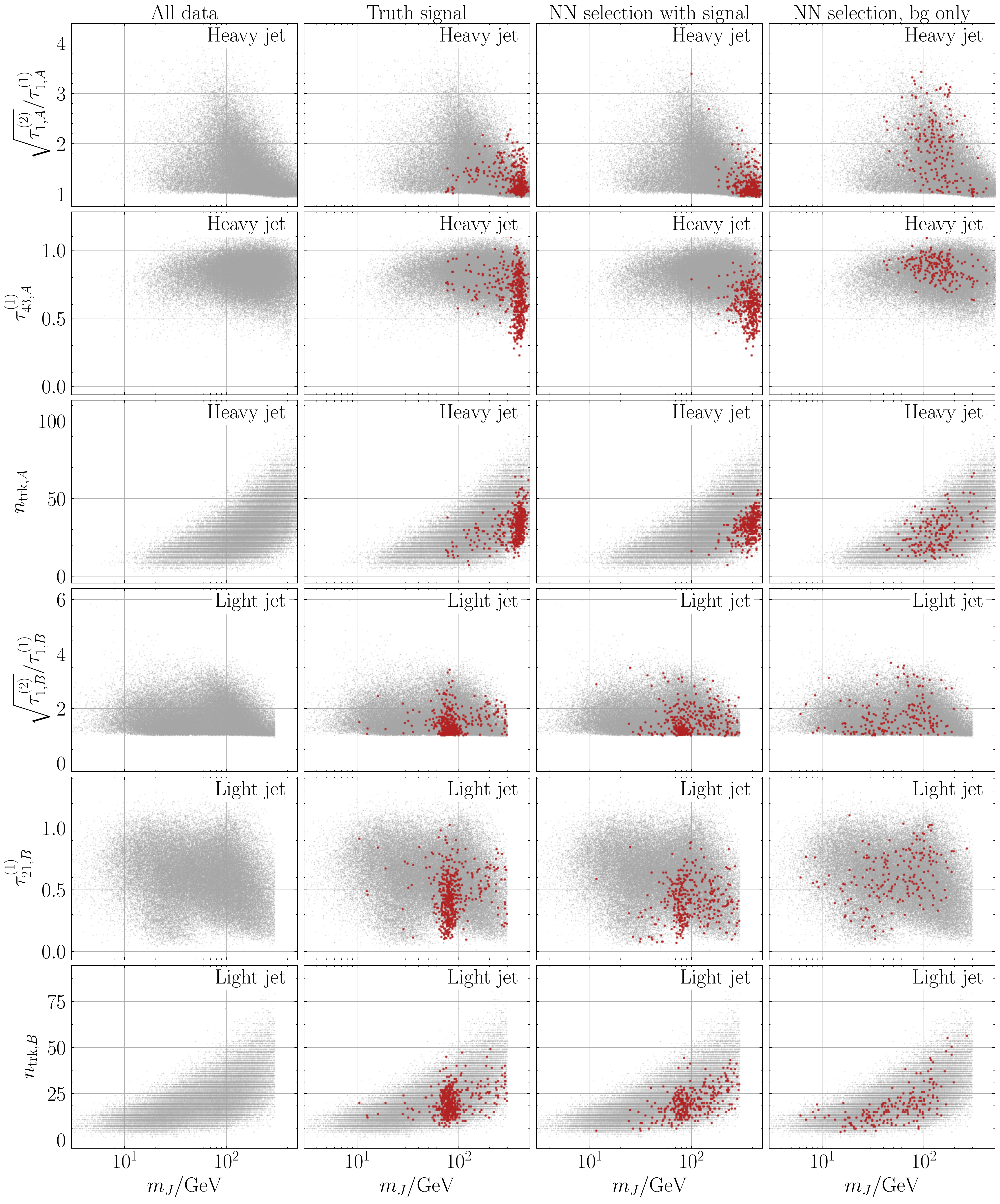}
\caption{2D projections of the 12D feature-space of the signal region dataset. \textbf{First column:} all signal region events. \textbf{Second column:} truth-level simulated signal events are highlighted in red. \textbf{Third column:} 0.2\% most signal-like events selected by the classifier desribed in Section~\ref{sec:physicsexample} are highlighted in red. \textbf{Fourth column:} highlighted in red are the 0.2\% most signal-like events selected by a classifier trained on the same sample but with true-signal events removed.}
\label{fig:scatterarray}
\end{figure}

Figure~\ref{fig:scatterarray} shows some further 2D projections of the data. In each case, the $x$-axis is the jet mass of the heavier or the lighter jet in the top three or bottom three rows, respectively, while the $y$-axes correspond to observables substructure variables as measured on the same jet. The first column are all events in the signal region. The second column are truth-level signal events in red overlaid on truth-level background in gray. The third column is the 0.2\% most signal-like events as determined by the classifier trained on this data. The fourth column shows the 0.2\% most signal-like events as determined by a classifier trained on the data sample with no signal data, only background. We see that the tagger trained when signal is present has found a cluster of events with a $400 \; \mathrm{GeV}$ jet with small $\tau_{43}^{(1)}$ and small $n_\text{trk}$; and an $80 \; \mathrm{GeV}$ jet with relatively small $\sqrt{\tau_1^{(2)}}/ \tau_1^{(1)}$, small $\tau_{21}^{(1)}$, and small $n_\text{trk}$. On the other hand, the events selected by the classifier trained on the background-only sample show no obvious clustering or pattern, and perhaps represent artifacts of statistical fluctuations in the training data.

\begin{figure}[!t]
\centering
\includegraphics[width=0.9\textwidth]{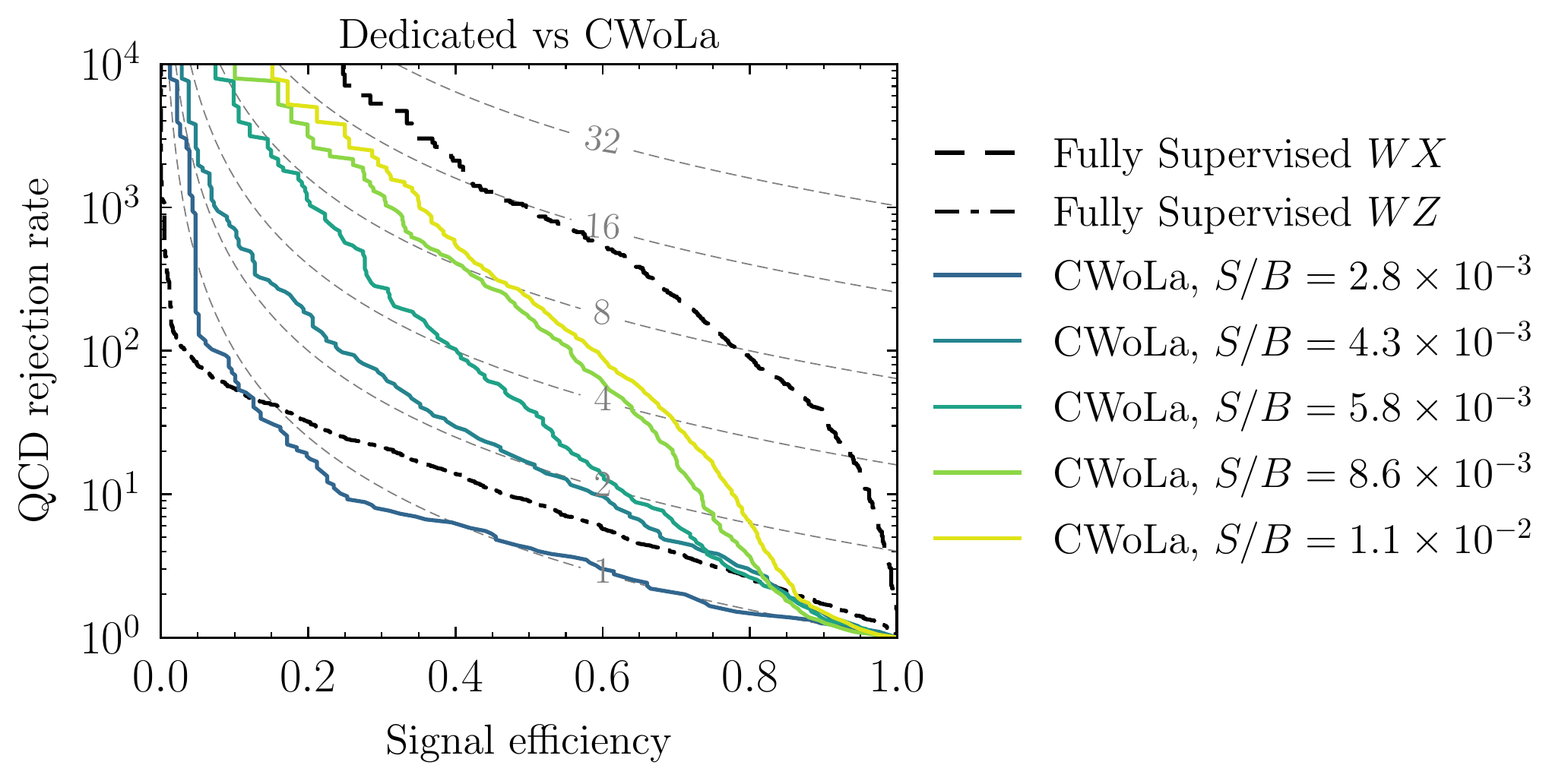}
\caption{Truth-label ROC curves for taggers trained using CWoLa with varying number of $WX$ signal events, compared to those for a dedicated tagger trained on pure $WX$ signal and background samples (dashed black) and one trained to discriminate $W$ and $Z$ jets from QCD (dot-dashed black). The CWoLa examples have $B = 81341$ in the signal region and $S = (230,352,472,697,927)$.}
\label{fig:dedvcwola}
\end{figure}

The ability of the CWoLa approach to discriminate signal from background depends on the number of signal and background events in the signal and sideband regions. In Fig.~\ref{fig:dedvcwola}, we keep the number of background events fixed but vary the size of the signal, and plot truth-label ROC curves for each example. This allows us to directly asses the performance of the taggers for the signal. For varying thresholds, the $x$-axis corresponds to the efficiency on true signal events in the signal region, $\epsilon_S$, while the $y$-axis represents the inverse efficiency on true QCD events in the signal region, $\epsilon_B$. The gray dashed lines labelled 1 to 32 indicate the significance improvement, $\epsilon_S / \sqrt{\epsilon_B}$, which quantifies the gain in statistical significance compared to the raw $m_{JJ}$ distribution with no cuts applied. In solid black we show the performance of a dedicated tagger trained with labelled signal and background events using a fully supervised approach. This gives a measure of the maximum achievable performance for this signal using the selected variables. A true dedicated tagger which could be used in a realistic dedicated search would be unlikely to reach this performance, since this would require careful calibration over 12 substructure variables with only simulated data available for the signal. While the CWoLa-based taggers do not reach the supervised performance in these examples, we find that performance does gradually improve with increasing statistics.

We also show in the dashed black curve the performance of $W$/$Z$ tagger in identifying this signal for which the tagger is not designed. This tagger is trained on a sample of $p p \to W' \to W Z$ events in the fully hadronic channel. In this case, the tagger is trained on the individual $W$/$Z$ jets themselves rather than the dijet event, as is typical in the current ATLAS and CMS searches. In producing the ROC curve, dijet events are considered to pass the tagging requirement if both large-radius jets pass a threshold cut on the output of the $W$/$Z$-tagger. We see that for $\epsilon_B \sim 10^{-4}$, which is a typical background rejection rate for recent hadronic diboson searches, the signal rate is negligible since the $X$-jet rarely passes the cuts. This illustrates that CWoLa hunting may find unexpected signals which are not targeted by existing dedicated searches is $S/B$ if high enough.  If $S/B$ is too low, then the CWoLa hunting approach is not able to identify the signal and it underperforms compared with the search that is targeting a different signal model. 

The datasets and code used for the case study can be found at Refs.~\cite{cwola_hunting_dataset, cwola_hunting_code}.

\section{Conclusions}
\label{sec:conc}

We have presented a new anomaly detection technique for finding BSM physics signals directly from data. The central assumption is that the signal is localized as a bump in one variable in which the background is smooth, and that other features are available for additional discrimination power.  This allows us to identify potential signal-enhanced and signal-depleted event samples with almost identical background characteristics on which a classifier can be trained using the Classification Without Labels approach. In the case that a distinctive signal is present, the trained classifier output becomes an effective discriminant between signal events and background events, while in the case that no signal is present the classifier output shows no clear pattern. An event selection based on a threshold cut on the classifier output produces a smooth distribution if no signal is present and produces a bump if a signal is present, and so standard bump hunting techniques can be used on the selected distribution.

The prototypical example used here is the dijet resonance search in which the dijet mass is the one-dimensional feature where the signal is localized.  Related quantities could also be used, such as the single jet mass for boosted resonance searches~\cite{Sirunyan:2017dnz,Sirunyan:2017nvi,Aaboud:2018zba} or the average mass of pair produced objects~\cite{Aaboud:2017nmi,CMS:2018sek,ATLAS:2012ds,Aad:2016kww,Chatrchyan:2013izb,Khachatryan:2014lpa}. Jet substructure information was used to augment information from just the dijet mass and a CWoLa classifier was trained using a deep neural network to discriminate signal region events from sideband events based on their substructure distributions.  Additional local information such as the number of leptons inside the jets, the number of displaced vertices, etc. could be used in the future to ensure sensitivity to a wide variety of models.  Furthermore, event-level information such as the number of jets or the magnitude of the missing transverse momentum could be added to an extended CWoLa hunt. 

The CWoLa hunting strategy is generalizable beyond this single case study. To summarize, the essential requirements are:
\begin{enumerate}
\item There is one bump-variable $m_\text{res}$ in which the background forms a smooth distribution, for which there is a background model such a parametric function, and a signal can be expected to be localized as a bump. This was the variable $m_{JJ}$ in the dijet case study.
\item There are additional features $Y$ in the events which may potentially provide discriminating power between signal and background, but the detailed topology of the of the signal in these variables is not known in advance. This was the set of substructure variables in the dijet study.
\item The background distribution in $Y$ should not have strong correlations with $m_\text{res}$ over the resonance width of the signal. In the case that such correlations exist, it may be possible to find a transformation of the variables that removes these correlations before being fed into the classifier, or alternatively to train the classifier in such a way that penalizes shaping of the $m_\text{res}$ distribution outside of the signal region.  Closure tests in simulation or with mixed samples in data can be used to confirm that $Y$ is not strongly correlated with $m_\text{res}$.
\end{enumerate}
By harnessing the power of modern machine learning, CWoLa hunting and other weakly supervised strategies may provide the key to uncovering BSM physics lurking in the unique datasets already collected by the LHC experiments.

\acknowledgments

We appreciate helpful discussions with and useful feedback on the manuscript from Timothy Cohen, Aviv Cukierman, Patrick Fox, Jack Kearney, Zhen Liu, Eric Metodiev, Brian Nord, Bryan Ostdiek, Matt Schwartz, and Jesse Thaler. We would also like to thank Peizhi Du for providing the UFO file for the benchmark signal model used in Sec.~\ref{sec:physicsexample}. The work of JHC is supported by NSF under Grant No. PHY-1620074 and by the Maryland Center for Fundamental Physics (MCFP). The work of B.N. is supported by the DOE under contract DE-AC02-05CH11231.
This manuscript has been authored by Fermi Research Alliance, LLC under
Contract No.  DE-AC02-07CH11359 with the U.S. Department of Energy,
Office of Science, Office of High Energy Physics. The United States
Government retains and the publisher, by accepting the article for
publication, acknowledges that the United States Government retains a
non-exclusive, paid-up, irrevocable, world-wide license to publish or
reproduce the published form of this manuscript, or allow others to do
so, for United States Government purposes.

\appendix

\section{Statistical Analysis}
\label{app:stats}
The significance of the bump is evaluated in the following way. First, after selecting the signal-like events in each cross-validation sample using its corresponding classifier, we merge the selected events from the $k$ samples into a single selected dataset. This dataset is binned in $m_{JJ}$, and we estimate the background by fitting a smooth, parametric function to the dataset with the signal region masked out. We use the following three-parameter function (also used in the ATLAS~\cite{Aaboud:2017eta} and CMS~\cite{Sirunyan:2016cao} searches for fully hadronic diboson resonances\footnote{More complex procedures for fitting the background such as Gaussian processes are also possible~\cite{Frate:2017mai} but their use is beyond our scope.}):

\begin{equation}
\label{equ:shape_function}
\frac{d N}{d m_{JJ}} = p_0 \frac{\left(1-m_{JJ} / \sqrt{s}\right)^{p_1}}{\left(m_{JJ} / \sqrt{s} \right)^{p_2}},
\end{equation}
which is fitted using a least-squares fit. The number of events in the signal region is predicted by summing the predictions in each of the three signal region bins. The systematic uncertainty in this fit is estimated by propagating linearly the uncertainties on the fit parameters onto an uncertainty in the signal region prediction. The fits and fit uncertainties are indicated in the left plot in Fig.~\ref{fig:pvalues} by the red dashed lines and gray bands. We tested the goodness of fit of this functional form in background-only simulations using Kolmogorov–Smirnov tests, and in any real search we would advocate similar tests in simulation. In the case that simulation is not completely reliable, it is possible to define data validation regions using non-signal selections in order to provide a cross-check of the fit function, as is done in e.g. Ref.~\cite{Aaboud:2017eta}. In the case of CWoLa hunting, this would entail selecting events in non-signal windows of the classifier output. For example, if using a 1\% selection for the signal search, one could use other percentile windows of the NN output to define non-signal selections with similar statistics which should be well fitted by the fit function under both the null and alternate hypotheses.

Since the shape of the signal in the signal region is a-priori unknown we base our hypothesis test on the total number of events in the signal region. We form the profile likelihood ratio
\begin{equation}
\lambda_0 = \frac{\mathcal{L}(\mu = 0, \doublehat{\theta})}{\mathcal{L}(\hat{\mu}, \hat{\theta})}
\end{equation}
where $\mu$ indicates the signal rate and $\theta$ is the nuisance parameter associated with the systematic uncertainties on the background prediction. In the numerator, $\doublehat{\theta}$ represents the best fit value for the nuisance parameter in the background-only hypothesis $\mu = 0$, while in the denominator $\hat{\mu}$ and $\hat{\theta}$ represent the combined best fit for $\mu$ and $\theta$. The likelihood is formed from a product of a Poisson factor for the number of events in the signal region, and a Gaussian constraint for the background nuisance parameter
\begin{equation}
\mathcal{L}(\mu, \theta) = \mathrm{Poiss}(n | b + \theta + \mu)  e^{- \theta^2/(2 \sigma^2)}
\end{equation}
where $n$ is the observed number of events in the signal region, $b$ is the number of background events predicted by the sideband fit, $\theta$ is the nuisance parameter associated with the systematic uncertainty for the background prediction, and $\sigma$ is the uncertainty on that nuisance parameter. 

Our test statistic is
\begin{equation}
q_0 = \begin{cases}
-2 \log(\lambda_0), &\hat{\mu} > 0,\\
0,  &\hat{\mu} \leq 0.
\end{cases}
\end{equation}
Using asymptotic formulae~\cite{Cowan:2010js}, gives a significance $Z = \sqrt{q_0}$ and $p_0 = 1 - \Phi(Z)$, where $\Phi$ is the cumulative distribution function of the normal distribution. 

The null hypothesis is that the dijet invariant mass distribution after selection by the classifiers is well described by the smooth functional form of Equation~(\ref{equ:shape_function}).  This requires that prior to any classification, the spectrum is smooth (already assumed by ATLAS and CMS) and that the classifiers are not able to generate localized features in the mass distribution following the CWoLa hunting procedure. In order to use the asymptotic formulae from Ref.~\cite{Cowan:2010js}, the bin counts in the selected, merged datasets must be Poissonian.  The rest of the section investigates the validity of these approximations.


Let $f(x)$ represent the function described by Eq.~\ref{equ:shape_function} prior to any classification and consider a dataset with $N_i^{\text{(uncut)}}$ events in $m_{JJ}$ bin $i$ (bin center $m_{\text{res}, \, i}$) with $N_i^{\text{(uncut)}}\sim\text{Poiss}(f(m_{\text{res}, \, i}))$.  Let $Y$ be a set of auxiliary observables whose probability distributions are independent of $m_\text{res}$.  The goal is to demonstrate that the $p$-values reported from the statistical procedure described above are accurate.  To begin, the dataset with $N_i^{\text{(uncut)}}$ events is partitioned into $k$ samples with equal probability for an event to be assigned to one of the samples.  Next, a classifier is trained to discriminate signal region events from sideband events using all subsamples except the $j$'th.  The classifier is then used to select a fraction $\epsilon$ of events in the held out $j$'th sample, using all other bins to determine $\epsilon$\footnote{The number of events used to determine $\epsilon$ is sufficiently large that the uncertainty on the value of the NN used to achieve $\epsilon$ efficiency is negligible.}.  This means that 

\begin{equation}
n_{j; \,i}^{\text{(cut)}} \sim \text{Poiss}\left( \frac{\epsilon}{k}f\left(m_{\text{res}, \, i}\right) \right).
\end{equation}


In the case that the cross-validated selected event rates in the $k$ samples are uncorrelated, then it would follow that after merging these datasets the total selected event rate distribution would be given by
\begin{equation}
N_i^{\text{(cut)}} = \sum_j n_{j; \,i}^{\text{(cut)}} \sim \text{Poiss}\left(\epsilon f\left(m_{\text{res}, \, i}\right) \right).
\end{equation}
However, because the events in one sample are used to train a classifier applied to the other samples, it cannot necessarily be assumed that the event rates are uncorrelated between samples. If strong correlations are expected between selected samples then in order to calculate reliable $p$-values the test statistic would need to be calibrated by running many toy experiments on either new simulated event samples or on bootstrapped samples, with the NNs trained fresh each time. Since this is a computationally expensive procedure, it is preferable if a simpler alternative is available.

\begin{figure}[h!]
\centering
\includegraphics[width=0.65\textwidth]{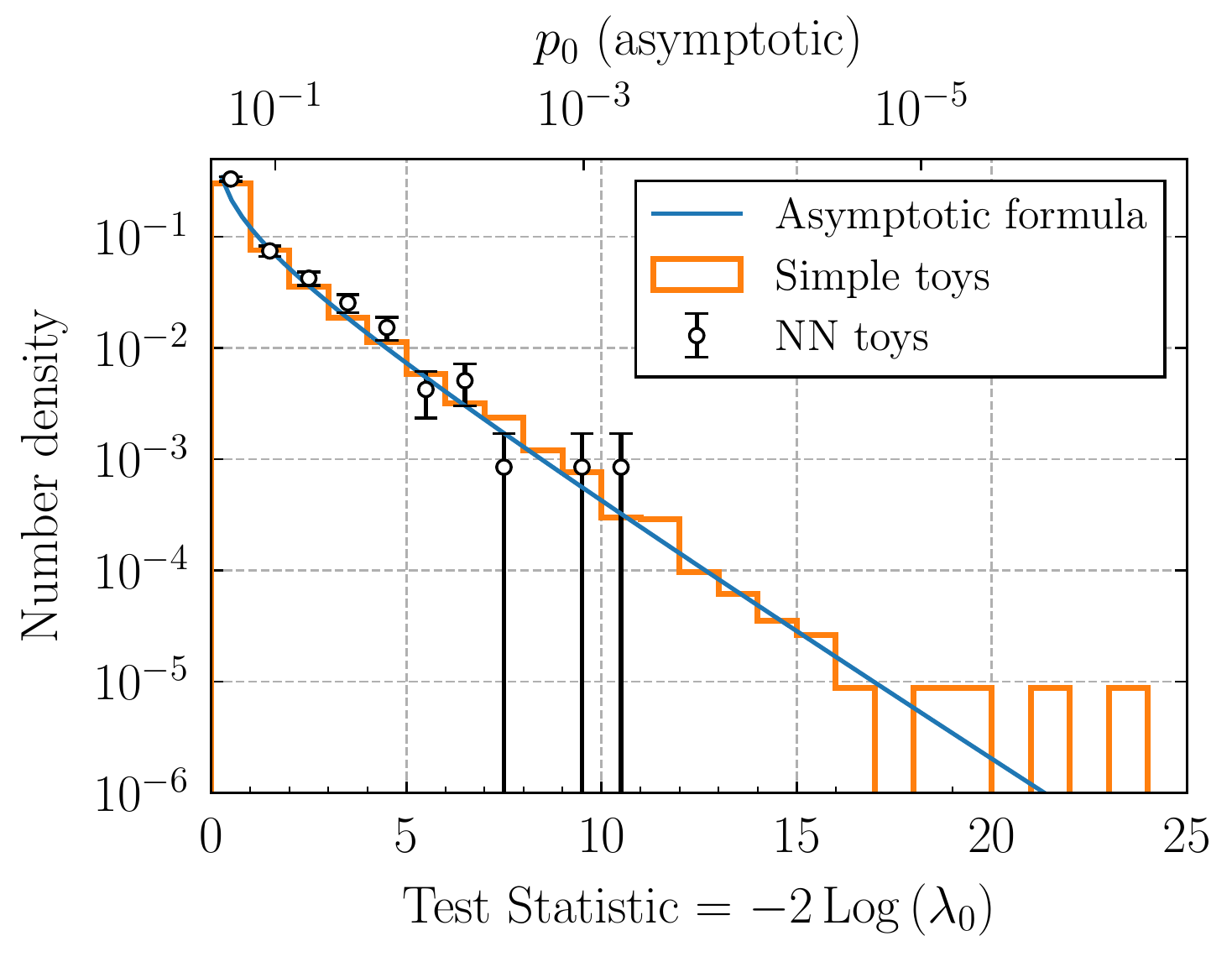}
\caption{Toy test statistic distributions. Black data points: $10^3$ toys with NN training and cross validation procedure; error bars represent $\sqrt{N}$ Poisson uncertainty. Orange histogram: $10^5$ toys with random selection (no NN training or cross-validation). Blue: Asymptotic formula.}
\label{fig:toy_stats}
\end{figure}

In order to check that the simpler approach (assuming no correlations between cross-validated samples) is valid, we have performed an empirical test of this effect in the following way. We generated $10^3$ toy datasets with binned event counts drawn from Poisson distributions with means determined by the distribution of Equation~(\ref{equ:shape_function}), with parameters obtained by a fit to the uncut dijet dataset used in Section.~\ref{sec:physicsexample}. Each event has 12 auxialliary variables, as in Section~\ref{sec:physicsexample}, but with these variables drawn from a random uniform distribution in the range $[0,1]$. NNs were trained using a cross validation procedure exactly as described in Section~\ref{sec:physicsexample}, except for the following modifications that were required to reduce the computational time required. We used 4-fold cross validation (rather than 5-fold), trained only four NNs per iteration from which the best was selected (rather than 20), and the NNs were trained with a patience of 100 epochs of no improvement in validation performance before stopping (instead of 300 epochs). The trained NNs were used to select the 1\% most `signal-like' events for each toy. For each toy we then calculated the test statistic for rejection of the null hypothesis, and the distribution of these test statistics is shown by the black markers in Figure~\ref{fig:toy_stats}.

Additionally, we generated $10^5$ toy datasets with $m_\text{res}$ drawn in the same way. Instead of training NNs to select events, we randomly selected 1\% of events. For each toy we then calculated the test statistic for rejection of the null hypothesis, and the distribution of these test statistics is shown by the orange histogram in Figure~\ref{fig:toy_stats}. Finally, we show the expected asymptotic distribution with the blue line.

The key feature of Figure~\ref{fig:toy_stats} is that the test statistic distribution for the NN toys shows no apparent deviation from that for the simple toys or from the asymptotic form. We therefore find no evidence of any distortion caused by correlations between cross-validation samples in this toy experiment.   

Finally, it is worth remarking that the $p$-value computed with the above procedure is only local.  If a local $p$-value is below some threshold, a followup, dedicated analysis using an orthogonal dataset should target the identified region of phase space with no trials factor penalty.  One could also estimate a global $p$-value in a standard way using e.g. a Bonferroni correction.  Other methods like the full bumphunter statistic could be used~\cite{Choudalakis:2011qn} but that is not the standard practice in the current ATLAS and CMS diboson resonance searches.

\section{Dijet Mass Scans}
\label{app:scan_plots}

\begin{figure}
\centering
\includegraphics[width=\textwidth]{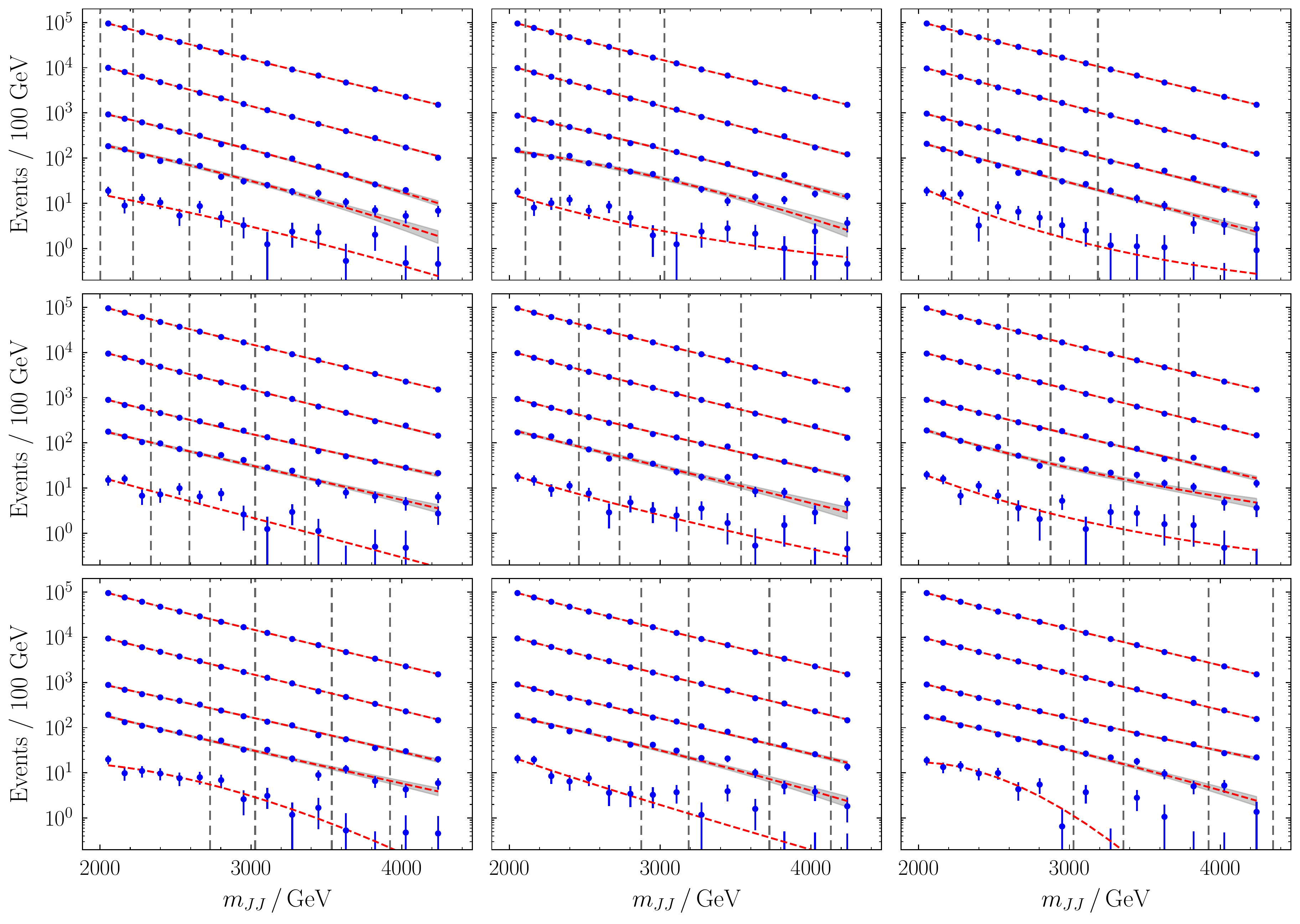}
\caption{Dijet invariant mass distributions in the resonance mass scan in the case that no signal is present. In each plot, the signal and sideband regions used for training the tagger are indicated by the vertical dashed lines. The top distribution in each plot is the original dijet mass distribution, and the subsequent distributions have had cuts applied at efficiencies of $10^{-1}, 10^{-2}, 2\times10^{-3}$, and $2\times10^{-4}$ respectively. The dashed red curves are fits determined by weighted least squares, and the gray bands the corresponding systematic uncertainty in the fit. In the lowest distribution of each plot the fit is poor as the presence of low-count bins makes the least squares fit inappropriate -- the fit line is kept to guide the eye, but the fit uncertainty band is ommitted.}
\label{fig:mJJarr-nosig}
\end{figure}

\begin{figure}
\centering
\includegraphics[width=\textwidth]{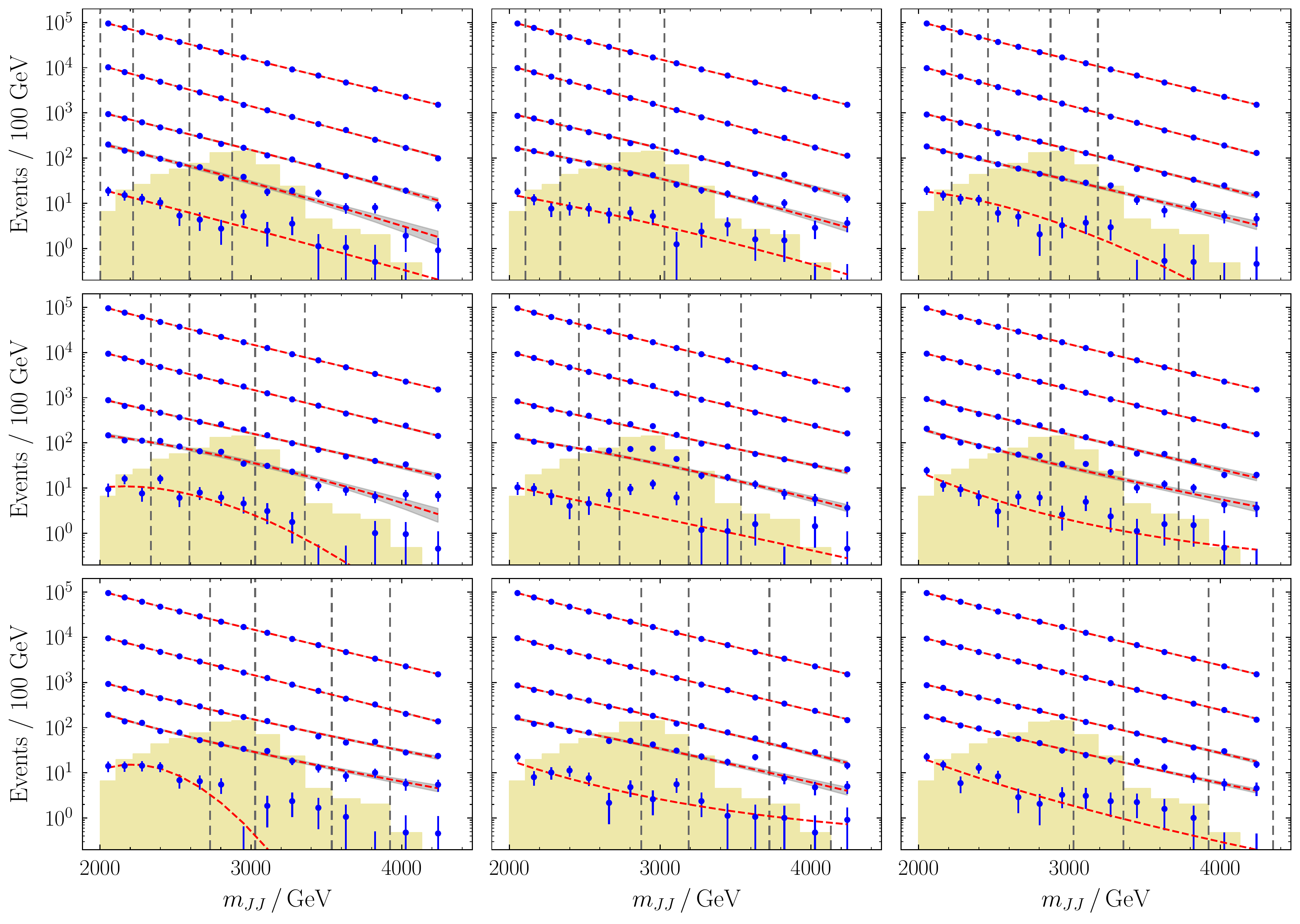}
\caption{Dijet invariant mass distributions in the resonance mass scan in the case that the signal is present (indicated by the filled histogram). Otherwise, the plots are as described in the caption to Fig.~\ref{fig:mJJarr-nosig}.}
\label{fig:mJJarr-sig}
\end{figure}

In Figs.~\ref{fig:mJJarr-nosig}, \ref{fig:mJJarr-sig} we plot the dijet invariant mass distributions before and after applying tagger cuts over the full range of the mass scan described in Sec.~\ref{sec:physicsexample}. The $p$-values calculated from the top four distributions in these plots are displayed in Fig.~\ref{fig:pvalues} (right).

\bibliographystyle{jhep}
\bibliography{myrefs}

\end{document}